\documentclass[amsmath,amssymb]{revtex4-2}

\usepackage{dcolumn}
\usepackage{hyperref}
\usepackage{chngcntr}
\usepackage{adjustbox}
\usepackage{amsmath}
\usepackage{siunitx}

\usepackage{graphicx}
\usepackage{amssymb}
\usepackage{epstopdf}
\usepackage{xcolor}
\usepackage{bm}
\usepackage{multirow}
\usepackage{subcaption}
\usepackage{tabularx}
\usepackage[figuresright]{rotating}
\usepackage{xcolor}
\usepackage{colortbl}
\usepackage{multirow}
\usepackage{ulem}

\makeatletter
\DeclareTextCompositeCommand{\r}{OT1}{A}{%
  \leavevmode\vbox{%
    \offinterlineskip
    \ialign{\hfil##\hfil\cr\char23\cr\noalign{\kern-1.15ex}A\cr}%
  }%
}
\makeatother

\renewcommand{\thefigure}{\Roman{figure}}
\begin{document}

\title{Significance of non-linear terms in the relativistic coupled-cluster theory in the determination of molecular properties}

\author{V. S. Prasannaa}
\email{srinivasaprasannaa@gmail.com}
\affiliation{Atomic, Molecular and Optical Physics Division, Physical Research Laboratory, Navrangpura, Ahmedabad 380009, India}

\author{B. K. Sahoo}
\email{bijaya@prl.res.in}
\affiliation{Atomic, Molecular and Optical Physics Division, Physical Research Laboratory, Navrangpura, Ahmedabad 380009, India}

\author{M. Abe}
\affiliation{Tokyo Metropolitan University, 1-1, Minami-Osawa, Hachioji City, Tokyo 192-0397, Japan}

\author{B. P. Das}
\affiliation{Department of Physics, Tokyo Institute of Technology, 2-12-1-H86 Ookayama, Meguro-ku, Tokyo 152-8550, Japan}

\begin{abstract}
\quad The relativistic coupled-cluster (RCC) theory has been applied recently to a number of heavy molecules to determine their properties very accurately. Since it demands large computational resources, the method is often approximated to singles and doubles excitations (RCCSD method). The effective electric fields (${\cal E}_{eff}$) and molecular permanent electric dipole moments (PDMs) of SrF, BaF and mercury monohalides (HgX with X = F, Cl, Br, and I) molecules are of immense interest for probing fundamental physics. In our earlier calculations of ${\cal E}_{eff}$ and PDMs for the above molecules, we had neglected the non-linear terms in the property evaluation expression of the RCCSD method. In this work, we demonstrate the roles of these terms in determining above quantities and their computational time scalability with number of processors of a computer. We also compare our results with previous calculations that employed variants of RCC theory as well as other many-body methods, and available experimental values. 
\end{abstract}

\maketitle

\section{Introduction}\label{sec1}

The coupled-cluster (CC) theory is considered to be the gold standard of electronic structure calculations in atoms and molecules~\cite{Kaldor, Nataraj}. It owes the title to its ability to capture electron correlation effects to a much better extent than other well-known many-body approaches such as configuration interaction (CI)~\cite{Bishop}, at a given level of truncation. This feature has led to accurate calculations of many properties in both the atomic and molecular systems (for example, see Refs.~\cite{BKS,MA}). We shall focus on the application of this method to evaluate molecular properties that are useful to probe fundamental physics, specifically the permanent electric dipole moment (PDM) and parity and time-reversal violating electric dipole moment of the electron (eEDM) \cite{Landau,Luders}. The molecular PDM is a very interesting property, and it plays a role in the sensitivity of an eEDM experiment through the polarizing factor \cite{HgX,FFCC}. The PDM is also an extremely relevant property in the ultracold sector, and molecules with large PDMs find  innumerable applications in that domain. For example, the SrF molecule possesses a fairly large PDM and hence gives rise to long-range, tunable, and anisotropic dipole-dipole interactions. This aspect, in combination with the fact that SrF is laser-coolable, makes the molecule important for applications such as exploring new quantum phases and quantum computing \cite{Shuman}. 

\begin{figure}[h!]
\centering
\includegraphics[width=13.00cm,height=7.0cm]{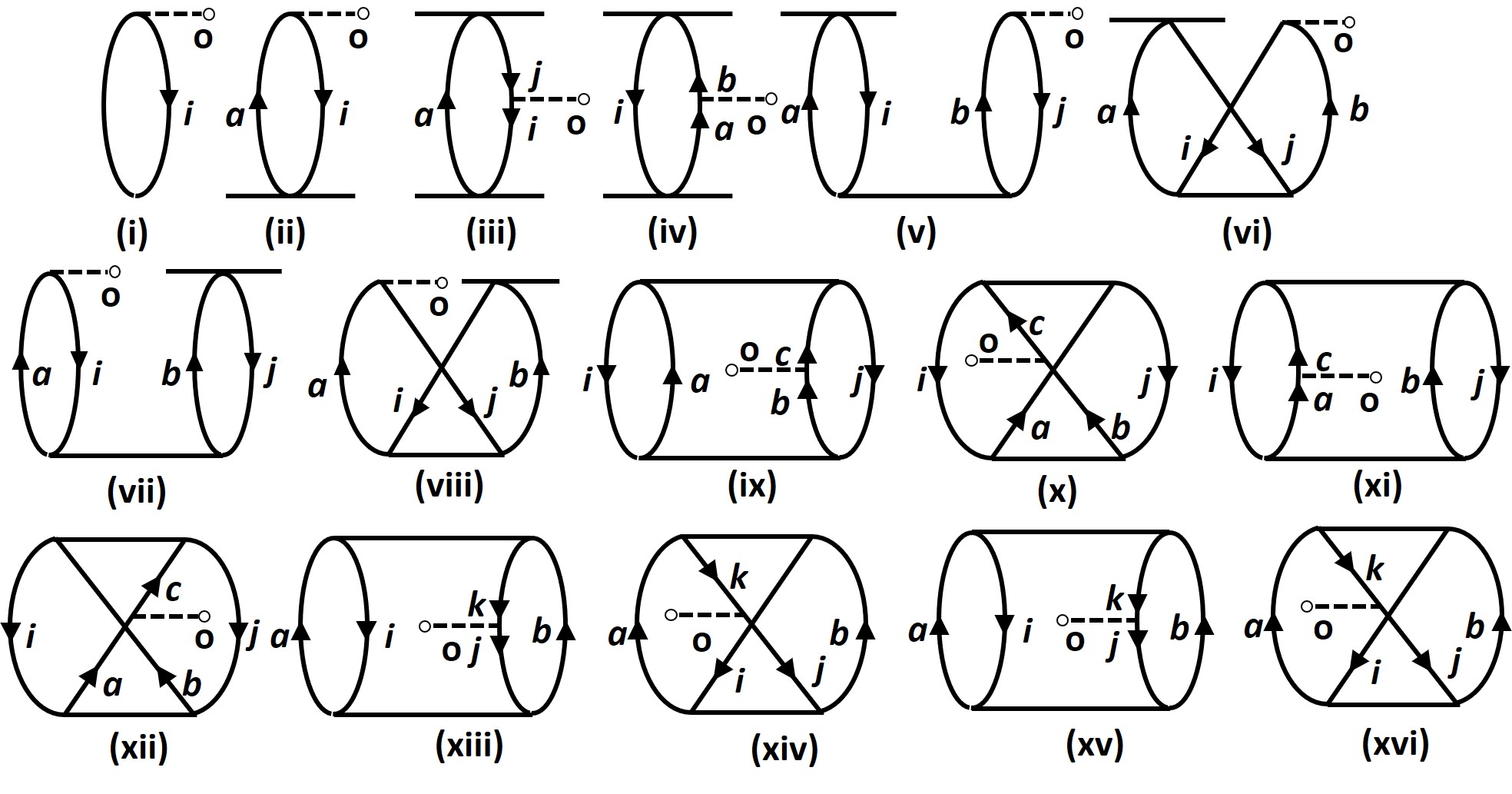}
\caption{Depictions of Goldstone diagrams representing the linear terms of the expectation value evaluation expression using the LERCCSD method. The notations $i, j, k, \cdots$ denote the hole lines, while $a, b, c ,\cdots$ denote the particle lines. Diagram (i) corresponds to contribution from the DF method, (ii)  is from the $OT_1$ term, (iii) and (iv) are from $T_1^\dag O T_1$, (v), (vi), (vii) and (viii) are diagrams for $T_1^\dag O T_2$, with (v) and (vii) corresponding to direct terms and (vi) and (viii) corresponding to exchange terms. Sub-figures (ix) to (xvi) include direct and exchange diagrams from  $T_2^\dag O T_2$. We also note that the hermitian conjugate diagrams of those given above are not explicitly sketched here. }
\label{fig:figure1}
\end{figure}

\begin{table}
\centering
\caption{\label{tab:table1} Contributions from the DF, LERCCSD and nLERCCSD methods to the $\mathcal{E}_\mathrm{eff}$s (in GV/cm) and PDMs (in Debye) of HgX, SrF, and BaF molecules from the present work (denoted as `This work' in the table). Comparison of the two properties from various works with our results are also presented.}
\begin{tabular}{llcc}
\hline
\hline
Molecule& Method & PDM&$\mathcal{E}_\mathrm{eff}$\\
\hline
SrF    & CASSCF-MRCI \cite{Jardali}      & 3.36&\\
    & CASSCF-RSPT2 \cite{Jardali}     & 3.61&\\
    & Z-vector \cite{Sasmal}          & 3.45&\\
    & LERCCSD \cite{AEM,FFCC}          & 3.6&2.17\\
    & FFCCSD \cite{FFCC}              & 3.62&2.16\\
    & X2C-MRCI \cite{Hao}             & 3.20&\\
    & X2C-FSCC \cite{Hao}             & 3.46&\\
    &DF ({\bf This work})             & 2.99&1.54\\
    & LERCCSD ({\bf This work})         & 3.57& 2.15\\
    & nLERCCSD ({\bf This work})        & 3.60 & 2.16 \\
    & Experiment~\cite{SrFexpt}       & 3.4676(1)& \\
BaF    & MRCI \cite{Tohme}               & 2.96&\\
    & LERCCSD \cite{AEM}                 & 3.4&6.50\\
    & FFCCSD \cite{FFCC}                & 3.41 & 6.46\\
    & X2C-MRCI \cite{Hao}             & 2.90 &\\
    & X2C-FSCC \cite{Hao}             & 3.23 &\\
    & Z-vector \cite{Talukdar}        & 3.08 &\\
    & ECP-RASSCF \cite{Kozlov}        &      & 7.5\\
    & RASCI \cite{Nayak}              &      &7.28\\
    & MRCI \cite{Meyer}               &      & 5.1\\
    & MRCI \cite{Meyer2}              &      & 6.1\\
    &DF ({\bf This work})             & 2.61&4.81\\
    & LERCCSD {\bf (This work)}       &3.32&6.45\\
    & nLERCCSD {\bf (This work)}         &   3.37  & 6.39 \\
    &Experiment (PDM)~\cite{BaFexpt}  & 3.17(3) &\\
HgF & CI \cite{Yu Yu}                  & 4.15    & 99.26\\
    & LERCCSD \cite{AJP}               &2.61     &\\
    & MRCI \cite{Meyer}               &         & 68\\
    & MRCI \cite{Meyer2}              &         & 95 \\
    &DF ({\bf This work})             & 4.11&105.69\\
    & LERCCSD \cite{HgX}              &         &115.42\\
    & FFCCSD \cite{FFCC}              & 2.92    & 116.37\\
    &  LERCCSD {\bf (This work)}       & 3.25    & 114.93\\
    &  nLERCCSD {\bf (This work)}      & 3.45    & 113.77 \\
HgCl & CI \cite{Wadt}            & 3.28    & \\
     & LERCCSD \cite{AJP}              & 2.72    & \\
    & LERCCSD \cite{HgX}               &         & 113.56 \\
    & FFCCSD \cite{FFCC}              & 2.96    & 114.31 \\
    &DF ({\bf This work})             & 4.30&104.33\\
    & LERCCSD {\bf (This work)}        & 3.26    & 112.51 \\
    & nLERCCSD {\bf (This work)}       & 3.45    & 110.94 \\
HgBr &  CI \cite{Wadt}                & 2.62    & \\
     & LERCCSD \cite{AJP}              & 2.36    & \\
     & LERCCSD \cite{HgX}              &         & 109.29 \\
     & FFCCSD \cite{FFCC}             & 2.71    & 109.56 \\
    &DF ({\bf This work})             & 4.14&99.72\\
     & LERCCSD {(\bf This work)}       & 2.62    & 109.38 \\
     & nLERCCSD {(\bf This work)}      & 2.94    & 107.42 \\
HgI  & LERCCSD \cite{AJP}              & 1.64    &\\
     & LERCCSD \cite{HgX}              &         & 109.3\\
     & FFCCSD \cite{FFCC}             & 2.06    & 109.56\\
    &DF ({\bf This work})             & 3.61&99.27\\
     & LERCCSD {(\bf This work)}       & 1.50    & 110.00\\
     & nLERCCSD {(\bf This work)}      & 2.01    & 107.38 \\
\hline \hline
\end{tabular}
\\
* The bond lengths chosen in out work are 2.00686 $A^{\circ}$, 2.42 $A^{\circ}$, 2.62 $A^{\circ}$, 2.81 $A^{\circ}$, 2.075 $A^{\circ}$, and 2.16 $A^{\circ}$ for HgF, HgCl, HgBr, HgI, SrF
and BaF, respectively. We used Dyall’s quadruple zeta (QZ) basis for Hg and I, Dunning’s correlation consistent polarized valence quadruple zeta (cc-pVQZ) basis for the halide atoms (F, Cl, and Br), and Dyall’s QZ functions augmented with Sapporo’s diffuse functions for Sr and Ba. 
\end{table}

\begin{sidewaystable}
\begin{center}
\caption{\label{tab:table2} Individual correlation contributions to the effective electric fields (in GV/cm) of mercury monohalides (HgX; X$=$F, Cl, Br, and I), SrF, and BaF, from the LERCCSD (abbreviated as `L') and nLERCCSD (denoted by `nL') methods. In the first column, the $A$ could be $O$ (which corresponds to LERCCSD diagrams) or $O_{x-y}$ (which is associated with nLERCCSD diagrams), where `x' and `y' could stand for the corresponding particle or hole line for a given term. The values are all rounded-off to two decimal places for HgX, while numbers that are extremely small in the case of SrF and BaF are denoted in the scientific notation instead. }
\begin{tabular}{l|l|cc|cc|cc|cc|cc|cc}
\hline
\hline
\multicolumn{2}{c|}{Molecule} & \multicolumn{2}{c|}{HgF} & \multicolumn{2}{c|}{HgCl}& \multicolumn{2}{c|}{HgBr}& \multicolumn{2}{c|}{HgI}& \multicolumn{2}{c|}{SrF}& \multicolumn{2}{c}{BaF} \\ \hline 
Term&Diagram&L&nL&L&nL&L&nL&L&nL&L&nL&L&nL\\ \hline 
&&\multicolumn{2}{c|}{}&\multicolumn{2}{c|}{}&\multicolumn{2}{c|}{}&\multicolumn{2}{c|}{}&\multicolumn{2}{c|}{}&\multicolumn{2}{c}{}\\
DF&Fig. 1(i)&\multicolumn{2}{c|}{105.69}&\multicolumn{2}{c|}{104.33}&\multicolumn{2}{c|}{99.72}&\multicolumn{2}{c|}{99.27}&\multicolumn{2}{c|}{1.54}&\multicolumn{2}{c}{4.81} \\ 
&&\multicolumn{2}{c|}{}&\multicolumn{2}{c|}{}&\multicolumn{2}{c|}{}&\multicolumn{2}{c|}{}&\multicolumn{2}{c|}{}&\multicolumn{2}{c}{}\\ \hline 
&&\multicolumn{2}{c|}{}&\multicolumn{2}{c|}{}&\multicolumn{2}{c|}{}&\multicolumn{2}{c|}{}&\multicolumn{2}{c|}{}&\multicolumn{2}{c}{}\\
$AT_1$&Fig. 1(ii)&17.09&13.11&17.05&12.21&19.83&14.76&23.85&15.71&0.63&0.61&1.79&1.60\\ 
&&\multicolumn{2}{c|}{}&\multicolumn{2}{c|}{}&\multicolumn{2}{c|}{}&\multicolumn{2}{c|}{}&\multicolumn{2}{c|}{}&\multicolumn{2}{c}{}\\ \hline 
&&\multicolumn{2}{c|}{}&\multicolumn{2}{c|}{}&\multicolumn{2}{c|}{}&\multicolumn{2}{c|}{}&\multicolumn{2}{c|}{}&\multicolumn{2}{c}{}\\
$T_1^\dag A T_1$&Fig. 1(iii)&$-$1.85&$-$0.28&$-$2.01&$-$0.25&$-$2.65&$-$0.62&$-$3.66&$-$0.41&$-$1.86$\times 10^{-2}$&$-$1.00$\times 10^{-3}$&$-$7.65$\times 10^{-2}$&$-$2.00$\times 10^{-4}$\\
&Fig. 1(iv)&$-$1.41&0.16&$-$1.40&0.28&$-$1.21&0.47&$-$1.56&1.16&$-$9.01$\times 10^{-3}$&4.80$\times 10^{-4}$&$-$6.47$\times 10^{-2}$&7.60$\times 10^{-3}$\\ 
&&\multicolumn{2}{c|}{}&\multicolumn{2}{c|}{}&\multicolumn{2}{c|}{}&\multicolumn{2}{c|}{}&\multicolumn{2}{c|}{}&\multicolumn{2}{c}{}\\ \hline 
&&\multicolumn{2}{c|}{}&\multicolumn{2}{c|}{}&\multicolumn{2}{c|}{}&\multicolumn{2}{c|}{}&\multicolumn{2}{c|}{}&\multicolumn{2}{c}{}\\
$T_1^\dag A T_2$&Fig. 1(v)&1.19&0.93&0.65&0.29&0.38&$-$0.11&0.38&$-$0.27&2.73$\times 10^{-3}$&1.02$\times 10^{-3}$&9.46$\times 10^{-3}$&2.51$\times 10^{-3}$\\ 
&Fig. 1(vi)&0.05&0.08&0.06&0.05&$-$0.01&$-$0.07&$-$0.03&$-$0.09&$-$4.91$\times 10^{-4}$&$-$7.93$\times 10^{-4}$&$-$1.49$\times 10^{-3}$&$-$2.39$\times 10^{-3}$\\ 
&Fig. 1(vii)&0.61&0.58&0.92&0.85&0.66&0.32&0.57&0.19&1.43$\times 10^{-2}$&1.48$\times 10^{-2}$&7.04$\times 10^{-2}$&7.13$\times 10^{-2}$\\ 
&Fig. 1(viii)&$-$1.31&$-$1.27&$-$1.24&$-$1.18&$-$0.91&$-$0.63&$-$1.26&$-$0.98&9.63$\times 10^{-3}$&9.91$\times 10^{-3}$&$-$2.49$\times 10^{-2}$&$-$2.32$\times 10^{-2}$\\ 
&&\multicolumn{2}{c|}{}&\multicolumn{2}{c|}{}&\multicolumn{2}{c|}{}&\multicolumn{2}{c|}{}&\multicolumn{2}{c|}{}&\multicolumn{2}{c}{}\\ \hline 
&&\multicolumn{2}{c|}{}&\multicolumn{2}{c|}{}&\multicolumn{2}{c|}{}&\multicolumn{2}{c|}{}&\multicolumn{2}{c|}{}&\multicolumn{2}{c}{}\\
$T_2^\dag A T_2$&Fig. 1(ix)&$-$2.50&$-$2.46&$-$2.54&$-$2.49&$-$2.68&$-$2.65&$-$2.93&$-$2.89&8.58$\times 10^{-3}$&6.15$\times 10^{-3}$&3.22$\times 10^{-2}$&2.17$\times 10^{-2}$\\ 
&Fig. 1(x)&$-$0.17&$-$0.17&$-$0.15&$-$0.14&$-$0.14&$-$0.13&$-$0.13&$-$0.11&$-$2.17$\times 10^{-3}$&$-$1.93$\times 10^{-3}$&$-$6.87$\times 10^{-3}$&$-$6.83$\times 10^{-3}$\\ 
&Fig. 1(xi)&$-$1.22&$-$1.40&$-$1.50&$-$1.47&$-$1.65&$-$1.85&$-$1.96&$-$1.99&$-$1.96$\times 10^{-2}$&$-$2.17$\times 10^{-2}$&$-$7.54$\times 10^{-2}$&$-$7.71$\times 10^{-2}$\\ 
&Fig. 1(xii)&$-$0.17&$-$0.17&$-$0.15&$-$0.14&$-$0.14&$-$0.13&$-$0.13&$-$0.11&$-$2.17$\times 10^{-3}$&$-$1.93$\times 10^{-3}$&$-$6.87$\times 10^{-3}$&$-$6.83$\times 10^{-3}$\\ 
&Fig. 1(xiii)&$-$1.64&$-$1.57&$-$1.67&$-$1.57&$-$1.70&$-$1.58&$-$1.84&$-$1.69&$-$1.38$\times 10^{-3}$&$-$1.96$\times 10^{-3}$&$-$3.20$\times 10^{-4}$&$-$9.61$\times 10^{-4}$\\ 
&Fig. 1(xiv)&$-$0.10&$-$0.10&$-$0.10&$-$0.10&$-$0.10&$-$0.10&$-$0.10&$-$0.10&$-$5.39$\times 10^{-4}$&$-$5.53$\times 10^{-4}$&$-$2.28$\times 10^{-3}$&$-$2.33$\times 10^{-3}$\\ 
&Fig. 1(xv)&0.77&0.74&0.36&0.37&0.08&0.12&$-$0.37&$-$0.21&4.42$\times 10^{-3}$&4.51$\times 10^{-3}$&2.82$\times 10^{-3}$&3.18$\times 10^{-3}$\\ 
&Fig. 1(xvi)&$-$0.10&$-$0.10&$-$0.10&$-$0.10&$-$0.10&$-$0.10&$-$0.10&$-$0.10&$-$5.39$\times 10^{-4}$&$-$5.53$\times 10^{-4}$&$-$2.28$\times 10^{-3}$&$-$2.33$\times 10^{-3}$\\ 
&&\multicolumn{2}{c|}{}&\multicolumn{2}{c|}{}&\multicolumn{2}{c|}{}&\multicolumn{2}{c|}{}&\multicolumn{2}{c|}{}&\multicolumn{2}{c}{}\\ \hline 
\multicolumn{2}{c|}{}&\multicolumn{2}{c|}{}&\multicolumn{2}{c|}{}&\multicolumn{2}{c|}{}&\multicolumn{2}{c|}{}&\multicolumn{2}{c|}{}&\multicolumn{2}{c}{}\\
\multicolumn{2}{c|}{Total}&114.93&113.77&112.51&110.94&109.38&107.42&110.00&107.38&2.15&2.16&6.45&6.39\\ 
\multicolumn{2}{c|}{}&\multicolumn{2}{c|}{}&\multicolumn{2}{c|}{}&\multicolumn{2}{c|}{}&\multicolumn{2}{c|}{}&\multicolumn{2}{c|}{}&\multicolumn{2}{c}{}\\ 
\hline \hline
\end{tabular}
\end{center}
\end{sidewaystable}

The extremely tiny eEDM is yet to be detected. Upper bound to it are extracted by a combination of relativistic many-body theory and experiment \cite{ACME2018}. These bounds, in turn, help to constrain several theories that lie beyond the Standard Model of particle physics, for example, supersymmetric theories \cite{Cesarotti}. A knowledge of the eEDM also aids in understanding the underlying physics that describes the matter-antimatter asymmetry in the universe~\cite{Fuyuto}. The theoretical molecular property of interest to eEDM is the effective electric field, $\mathcal{E}_\mathrm{eff}$. It is the internal electric field that is experienced by an electron due to other electrons and nuclei in a molecule. An accurate estimate of this quantity is used in setting or improving an upper bound to eEDM (for example, Ref. \cite{MA}), or to propose a new candidate for molecular eEDM experiments (for example, Ref. \cite{HgX}). This quantity can only be obtained using a \textit{relativistic} many-body theory~\cite{BPD}. Calculating the PDM provides information on polarizing factor for molecules that are proposed for an eEDM searches, where the property has not been measured. 

There have been several calculations of $\mathcal{E}_\mathrm{eff}$ for various molecules using the singles and doubles excitations approximation in the relativistic CC theory (RCCSD method), for example, Refs.~\cite{SPbF,SRaF}. In our earlier RCCSD calculations \cite{MA,HgX,Sunaga,HgA,RaH,YbOH}, the expectation value evaluating expression was approximated to only the linear terms (referred to as LERCCSD method).
Later, the calculations performed for HgX (X=F, Cl, Br, and I), SrF, and BaF besides other molecules were verified by using the finite-field energy derivative approach of the RCCSD theory (FFRCCSD method)~\cite{FFCC}, by adding the interaction Hamiltonians along with the residual Coulomb interaction operator. The LERCCSD and the FFRCCSD approaches showed excellent agreements (within 1 percent) in the values of $\mathcal{E}_\mathrm{eff}$. The results for the PDMs obtained in these methods were comparable for SrF and BaF and also were over-estimating the property with respect to their experimental values, but they differed substantially for HgX (with as much as 20 percent for HgI). The shortcomings of the above FFRCCSD method were that the accuracy of the results depended on numerical differentiation. Moreover, orbital relaxation effects were neglected by not including the perturbation in the Dirac-Fock (DF) level itself, in order to avoid breaking of Kramer's symmetry in the presence of a time-reversal symmetry violating eEDM interaction, which has to eventually be compensated for with further iterations. 

\begin{sidewaystable}
\begin{center}
\caption{\label{tab:table3} Correlation contributions to the PDMs (in Debye) of mercury monohalides (HgX; X$=$F, Cl, Br, and I), SrF, and BaF. The notation is same as in Table \ref{tab:table2}. The entry, `NC' stands for nuclear contribution to the PDM. }
\begin{tabular}{l|l|cc|cc|cc|cc|cc|cc}
\hline
\hline
\multicolumn{2}{c|}{Molecule} & \multicolumn{2}{c|}{HgF} & \multicolumn{2}{c|}{HgCl}& \multicolumn{2}{c|}{HgBr}& \multicolumn{2}{c|}{HgI}& \multicolumn{2}{c|}{SrF}& \multicolumn{2}{c}{BaF} \\ \hline
Term&Diagram&L&nL&L&nL&L&nL&L&nL&L&nL&L&nL\\ \hline
&&\multicolumn{2}{c|}{}&\multicolumn{2}{c|}{}&\multicolumn{2}{c|}{}&\multicolumn{2}{c|}{}&\multicolumn{2}{c|}{}&\multicolumn{2}{c}{}\\
DF&Fig. 1(i)&\multicolumn{2}{c|}{$-$767.04}&\multicolumn{2}{c|}{$-$925.61}&\multicolumn{2}{c|}{$-$1002.31}&\multicolumn{2}{c|}{$-$1075.83}&\multicolumn{2}{c|}{$-$375.75}&\multicolumn{2}{c}{$-$578.39}\\ 
&&\multicolumn{2}{c|}{}&\multicolumn{2}{c|}{}&\multicolumn{2}{c|}{}&\multicolumn{2}{c|}{}&\multicolumn{2}{c|}{}&\multicolumn{2}{c}{}\\ \hline
&&\multicolumn{2}{c|}{}&\multicolumn{2}{c|}{}&\multicolumn{2}{c|}{}&\multicolumn{2}{c|}{}&\multicolumn{2}{c|}{}&\multicolumn{2}{c}{}\\
$AT_1$&Fig. 1(ii)&$-$0.60&$-$0.78&$-$0.83&$-$1.01&$-$1.26&$-$1.54&$-$1.92&$-$2.33&0.63&0.65&0.80&0.82\\ 
&&\multicolumn{2}{c|}{}&\multicolumn{2}{c|}{}&\multicolumn{2}{c|}{}&\multicolumn{2}{c|}{}&\multicolumn{2}{c|}{}&\multicolumn{2}{c}{}\\ \hline
&&\multicolumn{2}{c|}{}&\multicolumn{2}{c|}{}&\multicolumn{2}{c|}{}&\multicolumn{2}{c|}{}&\multicolumn{2}{c|}{}&\multicolumn{2}{c}{}\\
$T_1^\dag A T_1$&Fig. 1(iii)&0.21&0.04&0.26&0.06&0.34&0.11&0.48&0.23&0.14&$-$0.01&0.19&$-$0.02\\ 
&Fig. 1(iv)&$-$0.45&0.05&$-$0.48&0.07&$-$0.62&0.13&$-$0.79&0.26&$-$0.18&$-$0.01&$-$0.23&$-$0.02\\ 
&&\multicolumn{2}{c|}{}&\multicolumn{2}{c|}{}&\multicolumn{2}{c|}{}&\multicolumn{2}{c|}{}&\multicolumn{2}{c|}{}&\multicolumn{2}{c}{}\\ \hline
&&\multicolumn{2}{c|}{}&\multicolumn{2}{c|}{}&\multicolumn{2}{c|}{}&\multicolumn{2}{c|}{}&\multicolumn{2}{c|}{}&\multicolumn{2}{c}{}\\
$T_1^\dag A T_2$&Fig. 1(v)&0.10&0.11&0.13&0.13&0.20&0.19&0.30&0.29&2.44$\times 10^{-2}$&2.53$\times 10^{-2}$&3.04$\times 10^{-2}$&3.13$\times 10^{-2}$\\ 
&Fig. 1(vi)&0.01&0.01&0.00&0.00&0.00&0.00&0.00&0.00&$-$1.99$\times 10^{-3}$&2.13$\times 10^{-3}$&2.40$\times 10^{-3}$&2.55$\times 10^{-3}$\\ 
&Fig. 1(vii)&0.01&0.01&0.01&$-$0.01&$-$0.01&$-$0.03&0.01&$-$0.01&9.48$\times 10^{-3}$&9.15$\times 10^{-3}$&9.46$\times 10^{-3}$&9.42$\times 10^{-3}$\\ 
&Fig. 1(viii)&0.02&0.01&0.01&0.01&0.02&0.03&0.04&0.04&1.47$\times 10^{-3}$&1.45$\times 10^{-3}$&$-$4.02$\times 10^{-3}$&4.10$\times 10^{-3}$\\ 
&&\multicolumn{2}{c|}{}&\multicolumn{2}{c|}{}&\multicolumn{2}{c|}{}&\multicolumn{2}{c|}{}&\multicolumn{2}{c|}{}&\multicolumn{2}{c}{}\\ \hline
&&\multicolumn{2}{c|}{}&\multicolumn{2}{c|}{}&\multicolumn{2}{c|}{}&\multicolumn{2}{c|}{}&\multicolumn{2}{c|}{}&\multicolumn{2}{c}{}\\
$T_2^\dag A T_2$&Fig. 1(ix)&1.19&1.19&1.48&1.47&1.66&1.66&1.84&1.85&0.82&0.82&0.98&0.97\\ 
&Fig. 1(x)&$-$0.01&$-$0.01&0.00&0.00&0.00&0.00&0.00&0.00&9.92$\times 10^{-4}$&1.57$\times 10^{-3}$&$-$2.43$\times 10^{-3}$&$-$2.99$\times 10^{-3}$\\  
&Fig. 1(xi)&1.14&1.16&1.40&1.41&1.57&1.62&1.73&1.81&0.79&0.78&0.95&0.94\\ 
&Fig. 1(xii)&$-$0.01&$-$0.01&0.00&0.00&0.00&0.00&0.00&0.00&9.22$\times 10^{-4}$&1.57$\times 10^{-3}$&$-$2.43$\times 10^{-3}$&$-$2.99$\times 10^{-3}$\\  
&Fig. 1(xiii)&$-$1.26&$-$1.25&$-$1.53&$-$1.52&$-$1.72&$-$1.70&$-$1.91&$-$1.89&$-$0.84&$-$0.84&$-$1.01&$-$1.01\\  
&Fig. 1(xiv)&0.01&0.01&0.01&0.01&0.00&0.00&0.01&0.00&7.35$\times 10^{-3}$&7.37$\times 10^{-3}$&$-$8.38$\times 10^{-3}$&0.01\\  
&Fig. 1(xv)&$-$1.23&$-$1.21&$-$1.51&$-$1.48&$-$1.70&$-$1.67&$-$1.91&$-$1.85&$-$0.83&$-$0.83&$-$0.99&$-$0.98\\ 
&Fig. 1(xvi)&0.01&0.01&0.01&0.01&0.00&0.00&0.01&0.00&7.35$\times 10^{-3}$&7.37$\times 10^{-3}$&$-$8.38$\times 10^{-3}$&0.01\\ 
&&\multicolumn{2}{c|}{}&\multicolumn{2}{c|}{}&\multicolumn{2}{c|}{}&\multicolumn{2}{c|}{}&\multicolumn{2}{c|}{}&\multicolumn{2}{c}{}\\ \hline
\multicolumn{2}{c|}{}&\multicolumn{2}{c|}{}&\multicolumn{2}{c|}{}&\multicolumn{2}{c|}{}&\multicolumn{2}{c|}{}&\multicolumn{2}{c|}{}&\multicolumn{2}{c}{}\\
\multicolumn{2}{c|}{NC}&\multicolumn{2}{c|}{771.15}&\multicolumn{2}{c|}{929.91}&\multicolumn{2}{c|}{1006.45}&\multicolumn{2}{c|}{1079.44}&\multicolumn{2}{c|}{378.74}&\multicolumn{2}{c}{581.00}\\ 
\multicolumn{2}{c|}{}&\multicolumn{2}{c|}{}&\multicolumn{2}{c|}{}&\multicolumn{2}{c|}{}&\multicolumn{2}{c|}{}&\multicolumn{2}{c|}{}&\multicolumn{2}{c}{}\\ \hline
\multicolumn{2}{c|}{}&\multicolumn{2}{c|}{}&\multicolumn{2}{c|}{}&\multicolumn{2}{c|}{}&\multicolumn{2}{c|}{}&\multicolumn{2}{c|}{}&\multicolumn{2}{c}{}\\
\multicolumn{2}{c|}{Total}&3.25&3.45&3.26&3.45&2.62&2.94&1.50&2.01&3.57&3.60&3.32&3.37\\ 
\multicolumn{2}{c|}{}&\multicolumn{2}{c|}{}&\multicolumn{2}{c|}{}&\multicolumn{2}{c|}{}&\multicolumn{2}{c|}{}&\multicolumn{2}{c|}{}&\multicolumn{2}{c}{}\\
\hline \hline
\end{tabular}
\end{center}
\end{sidewaystable}

Here, we intend to calculate the values of ${\cal E}_{eff}$ and PDM by including the non-linear terms in the expectation value evaluation expression of the RCCSD method (nLERCCSD method). We adopt the intermediate-diagram approach as discussed in Refs.~\cite{Bartlett,Yash} to implement these non-linear RCC terms. For this purpose, we have undertaken molecules that are very relevant for eEDM studies. HgX molecules were identified as promising candidates for future eEDM searches, owing to their extremely large effective electric fields as well as experimental advantages~\cite{HgX}. A recent work that proposes to laser-cool HgF has opened new avenues for an upcoming eEDM experiment with the molecule~\cite{HgFlc}. Another very important molecule in this regard is BaF, and two eEDM experiments are simultaneously underway for this system~\cite{15,15prime}. Experimental values of the PDMs are available only for BaF among the systems that we mention above. We also present results for the PDM of SrF as it was the first molecule to be laser-cooled~\cite{Shuman}, and a very precise measurement of this quantity has been reported~\cite{SrFexpt}. 

\section{Theory and Implementation}\label{sec2}

In the RCC theory, the wave function of a molecular state  is expressed as \cite{Cizek}
\begin{eqnarray}
 \arrowvert \Psi \rangle = e^{T} \arrowvert \Phi_{0} \rangle, 
\end{eqnarray}
where $T$ is the cluster operator and $\arrowvert \Phi_{0} \rangle$ is the reference state obtained by mean-field theory. We use the Dirac-Coulomb Hamiltonian in our calculations, and $\arrowvert \Phi_0 \rangle$ is obtained using the DF method. In the RCCSD method, we approximate $T = T_1 + T_2$ with subscripts $1$ and $2$ indicating singles and doubles excitations, respectively, and they are given using the second-quantization operators as 
\begin{eqnarray}
 T_1 &=& \sum_{i,a} t_i^a a_a^{\dag}a_i \ \ \ \text{and} \ \ \ 
 T_2 = \frac{1}{2} \sum_{i,j,a,b} t_{ij}^{ab} a_a^{\dag}a_b^{\dag}a_j a_i ,
\end{eqnarray}
where the notation $i, j$ is used to denote holes $a, b$ refer to particles, $t_i^a$ is the one hole-one particle excitation amplitude and $t_{ij}^{ab}$ is the two-hole two-particle excitation amplitude. 

We have employed the UTChem~\cite{utchem,utchem2} program for the DF calculations, the atomic orbital to molecular orbital integral transformations as well as generating the property integrals, and the Dirac08~\cite{dirac} code to obtain the RCCSD excitation operator amplitudes. It is important to reiterate that all the non-linear terms were included in the equations of the RCCSD method to determine the excitation amplitudes. 

\begin{figure*}[t]
\centering
\begin{tabular}{cccccc}
\includegraphics[width=1.3cm,height=1.4cm]{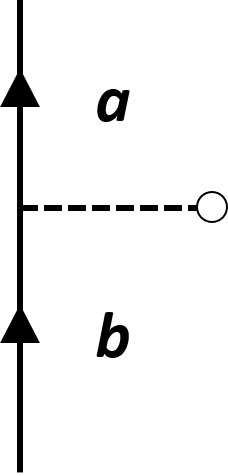} & \includegraphics[width=1.6cm,height=1.4cm]{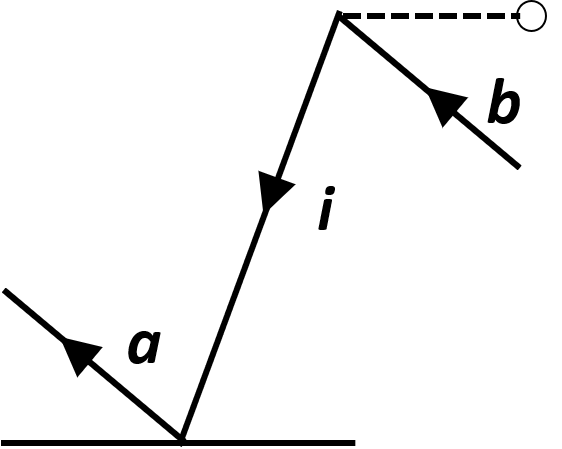} & \includegraphics[width=1.6cm,height=1.4cm]{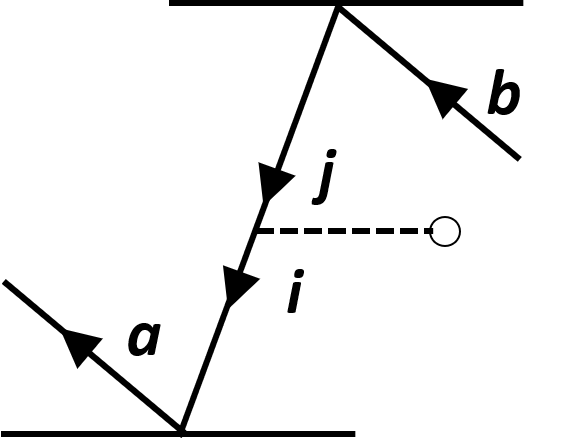} &
\includegraphics[width=1.6cm,height=1.4cm]{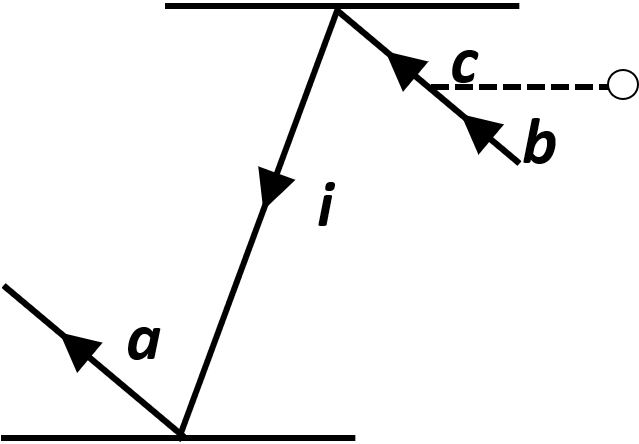} &
\includegraphics[width=1.6cm,height=1.4cm]{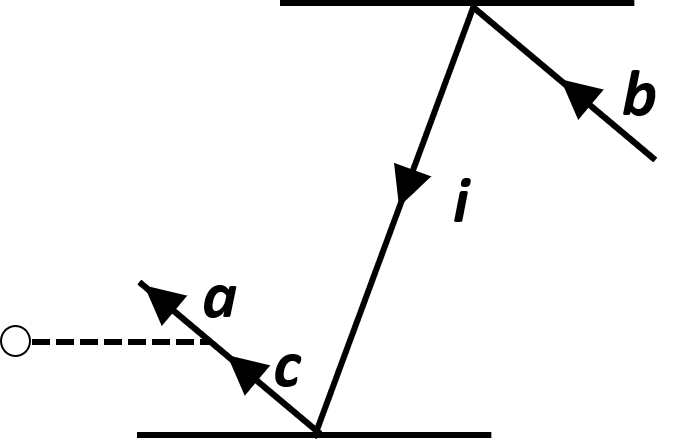} &
\includegraphics[width=2.2cm,height=1.4cm]{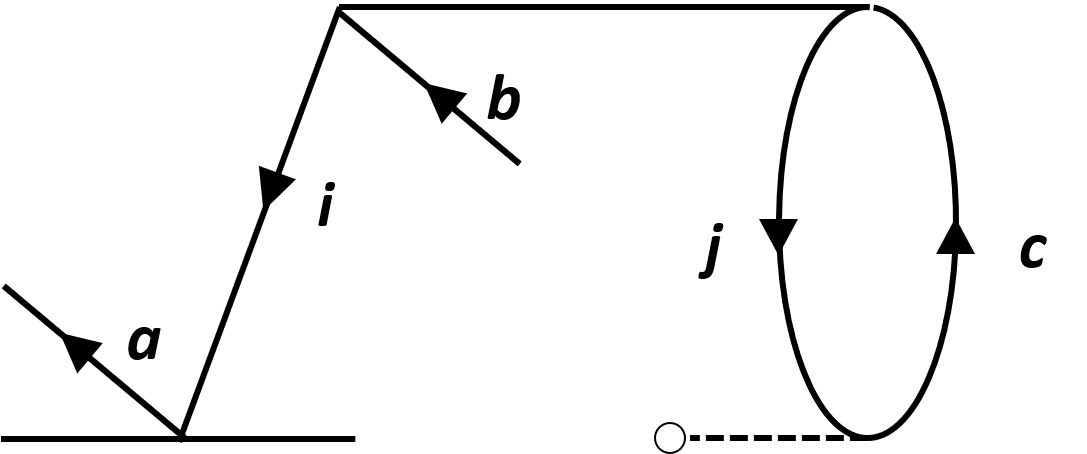}\\
(i) & (ii) & (iii) & (iv) & (v) & (vi) \\ \\
\includegraphics[width=2.2cm,height=1.4cm]{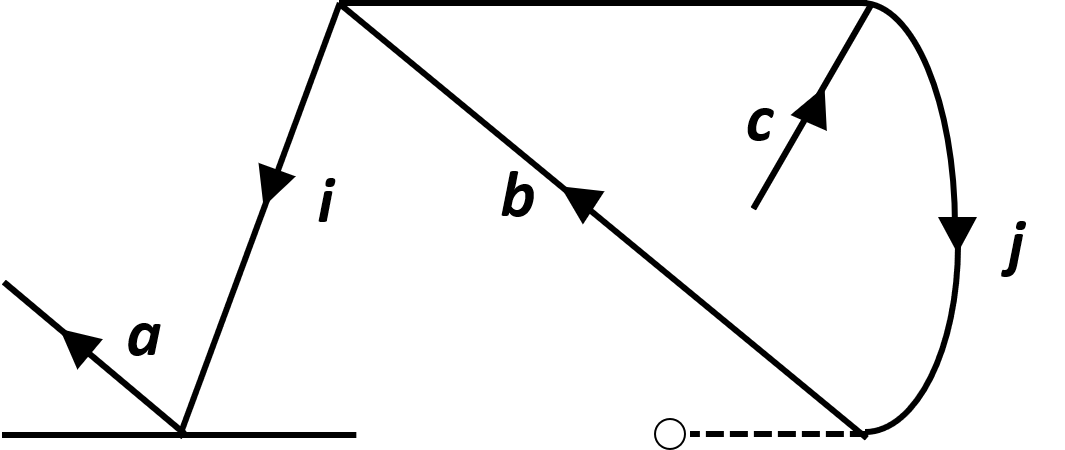} & \includegraphics[width=2.2cm,height=1.4cm]{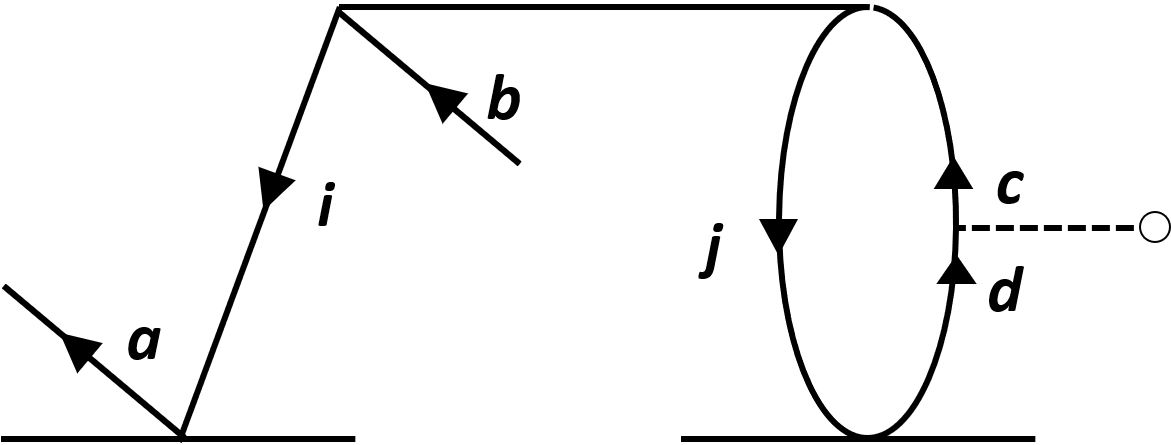} & \includegraphics[width=2.2cm,height=1.4cm]{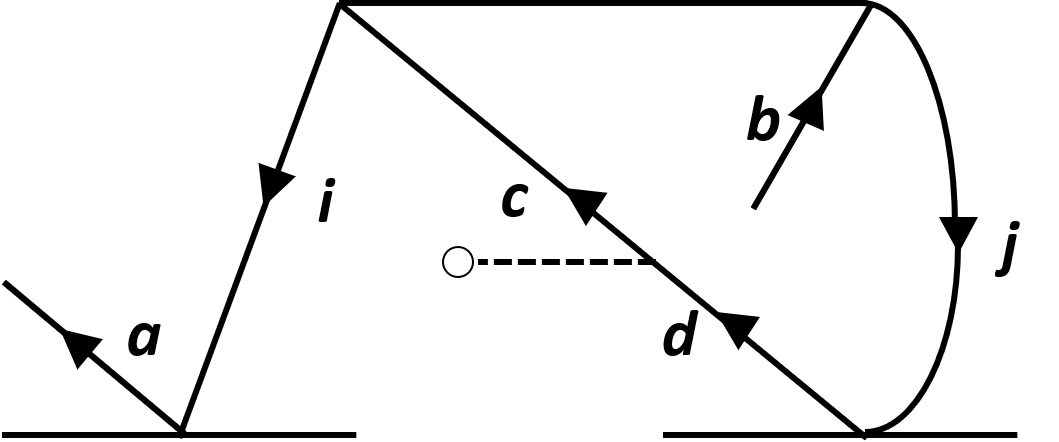} &
\includegraphics[width=2.2cm,height=1.4cm]{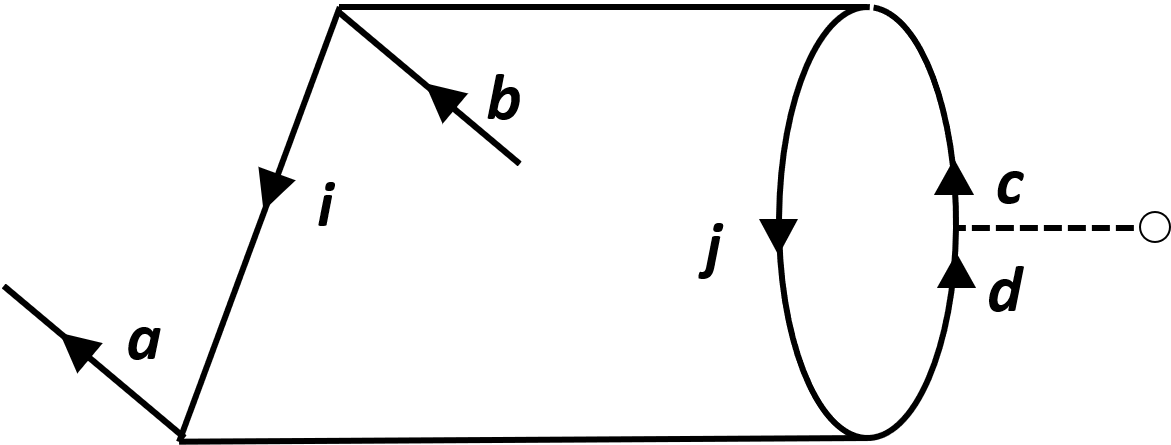} &
\includegraphics[width=2.2cm,height=1.4cm]{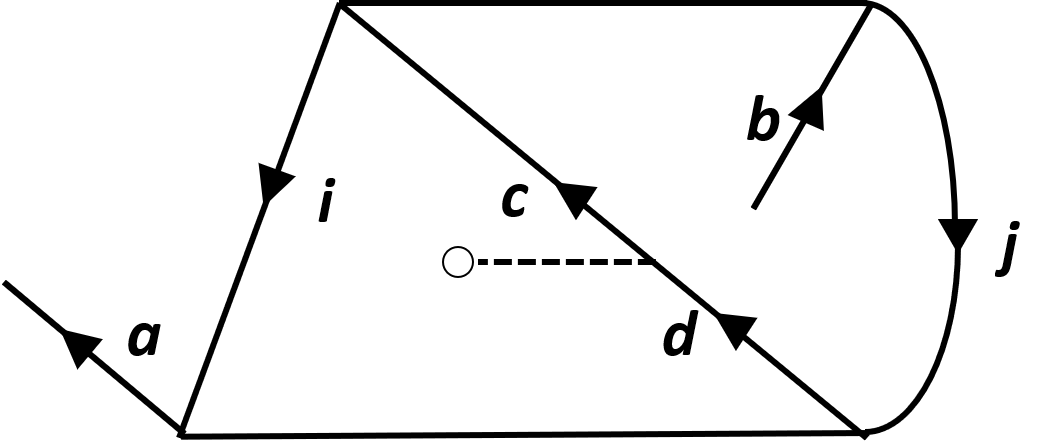} &
\includegraphics[width=2.2cm,height=1.4cm]{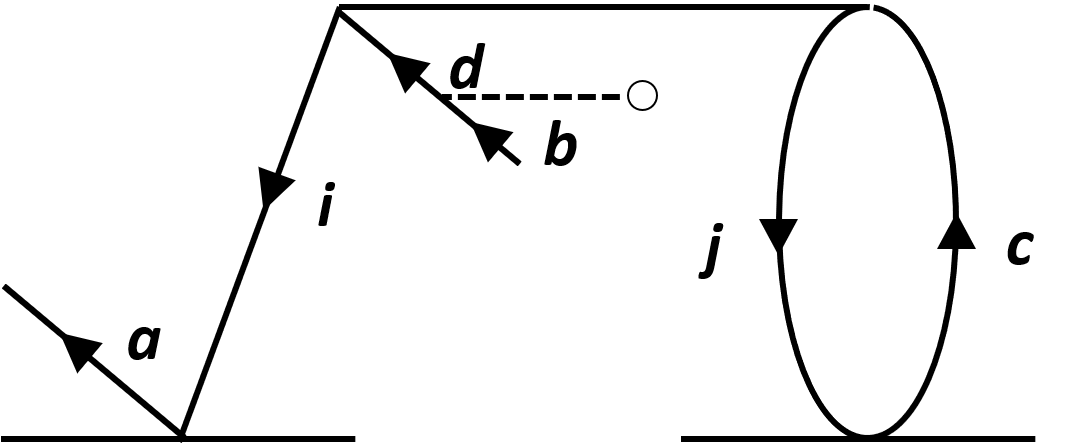}\\
(vii) & (viii) & (ix) & (x) & (xi) & (xii)\\ \\
\includegraphics[width=2.2cm,height=1.4cm]{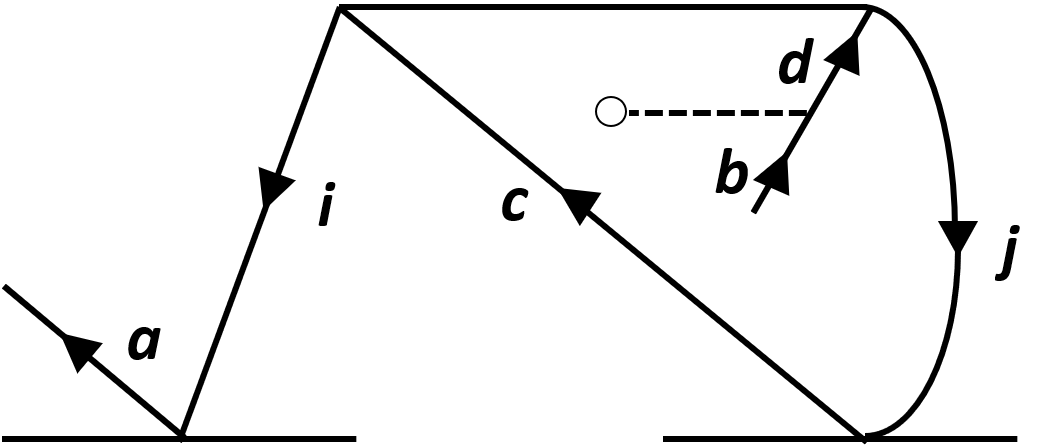} & \includegraphics[width=2.2cm,height=1.4cm]{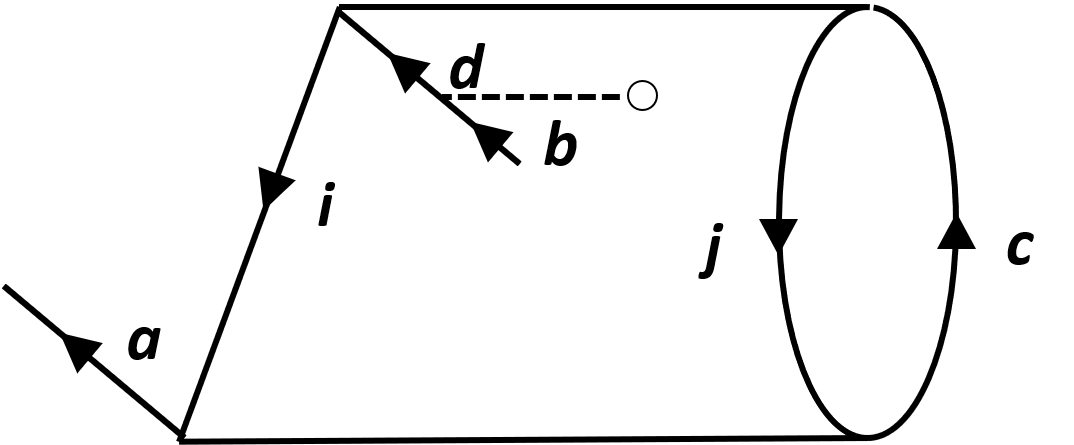} & \includegraphics[width=2.2cm,height=1.4cm]{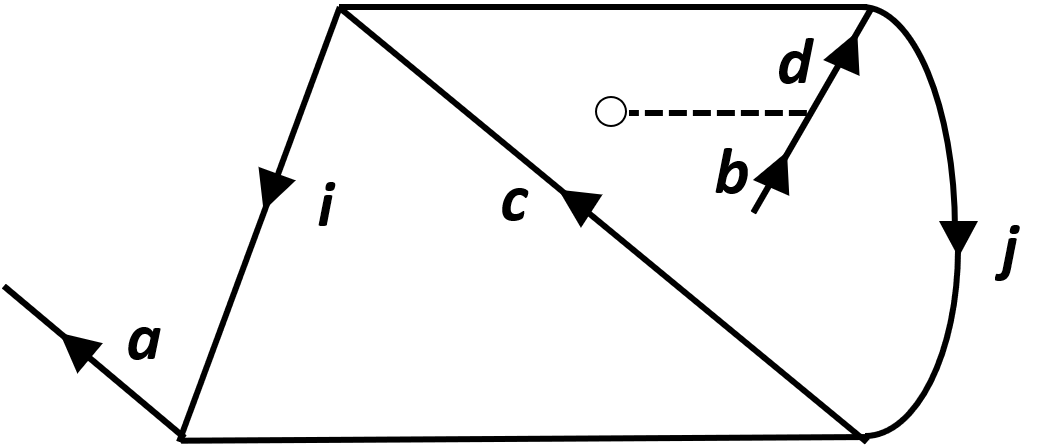} &
\includegraphics[width=2.2cm,height=1.4cm]{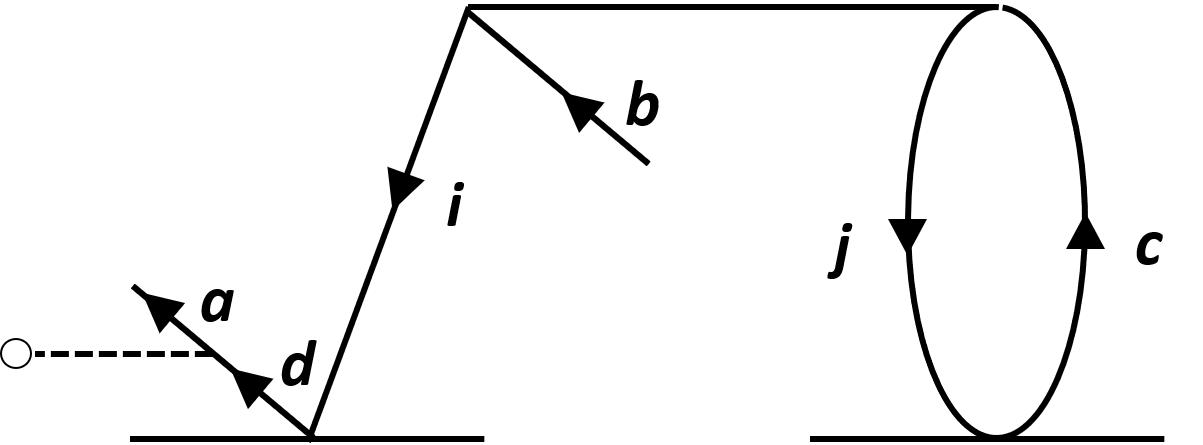} &
\includegraphics[width=2.2cm,height=1.4cm]{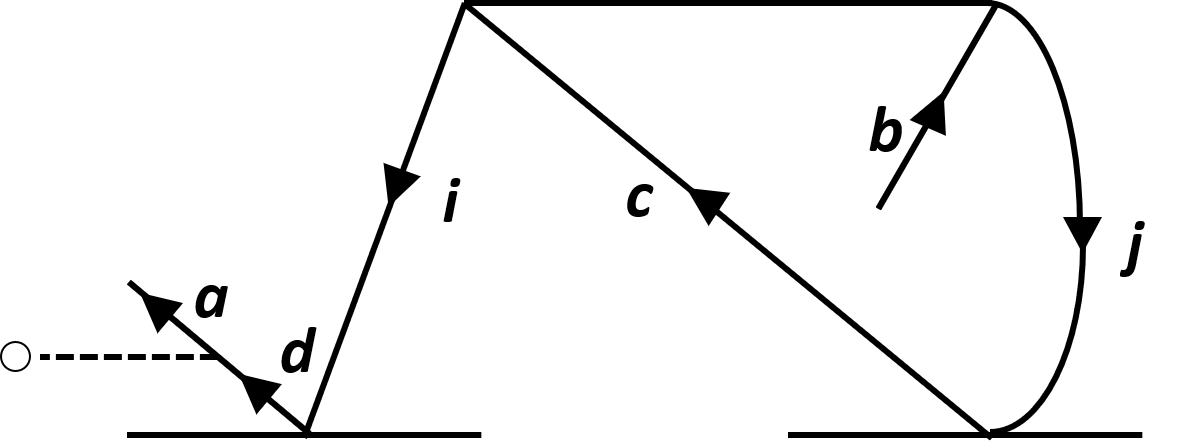} &
\includegraphics[width=2.2cm,height=1.4cm]{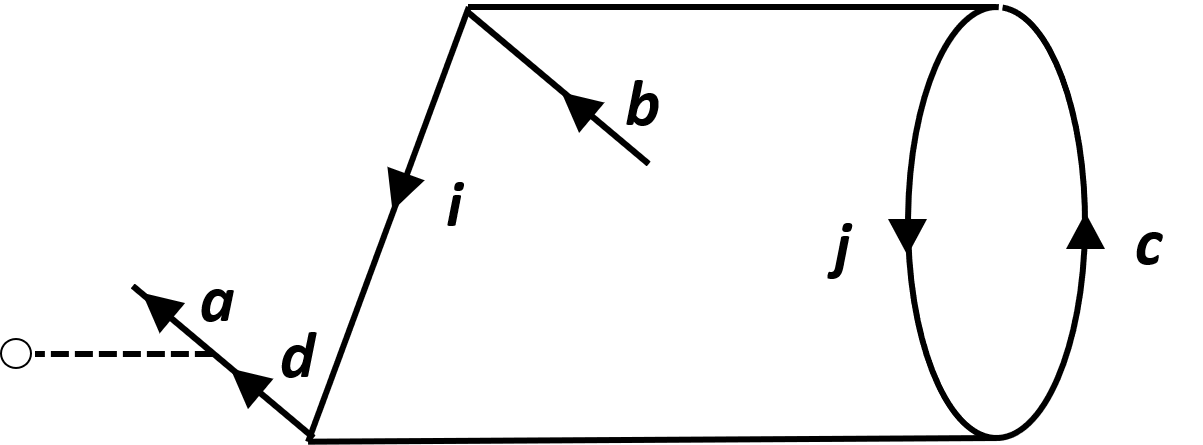}\\
(xiii) & (xiv) & (xv) & (xvi) & (xvii) & (xviii)\\ \\
\includegraphics[width=2.2cm,height=1.4cm]{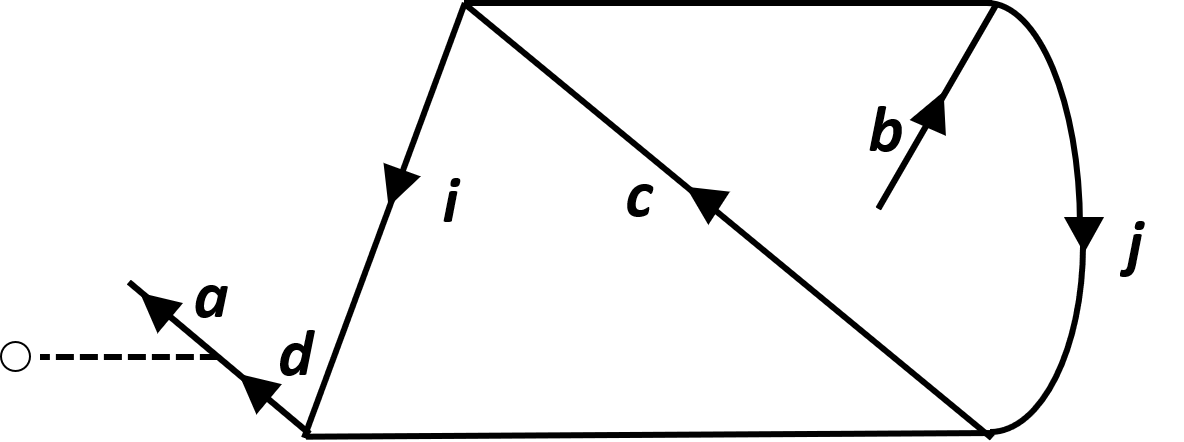} & \includegraphics[width=2.2cm,height=1.4cm]{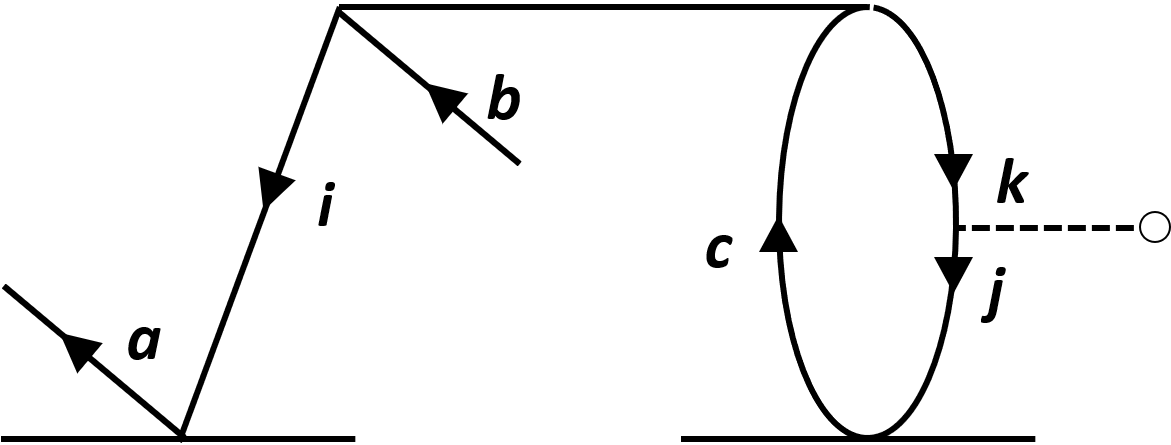} & \includegraphics[width=2.2cm,height=1.4cm]{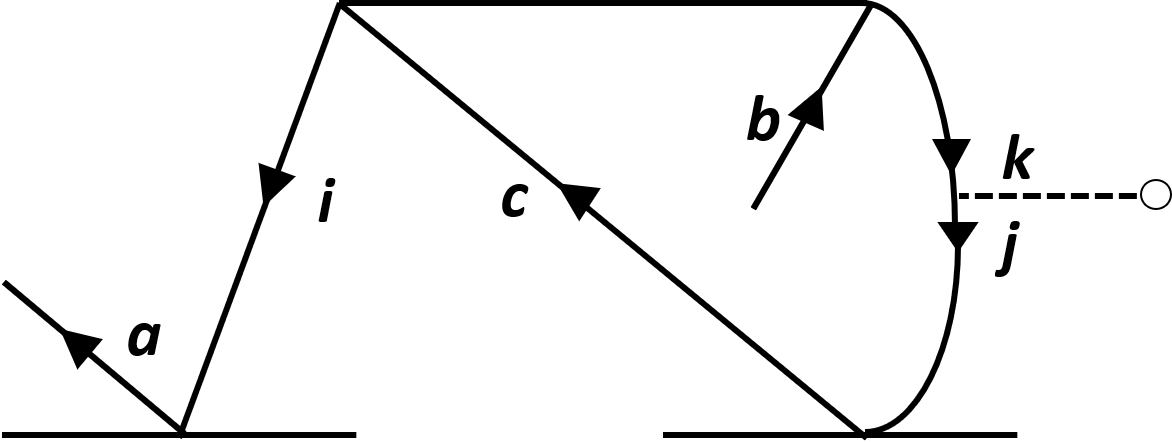} &
\includegraphics[width=2.2cm,height=1.4cm]{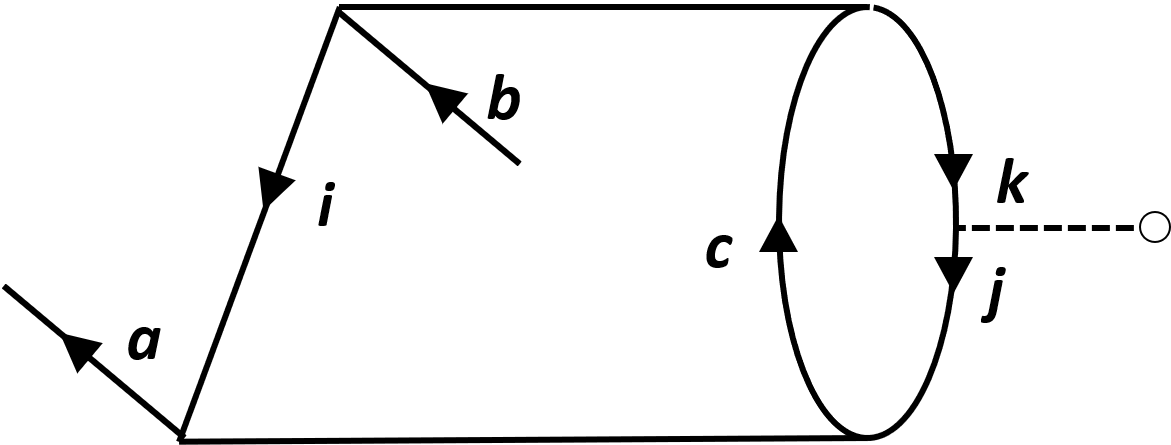} &
\includegraphics[width=2.2cm,height=1.4cm]{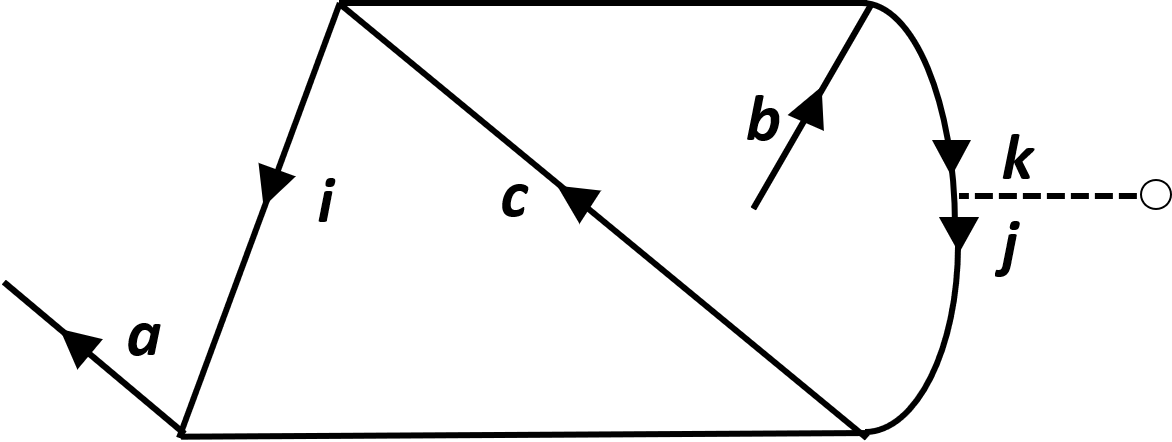} &
\includegraphics[width=2.2cm,height=1.4cm]{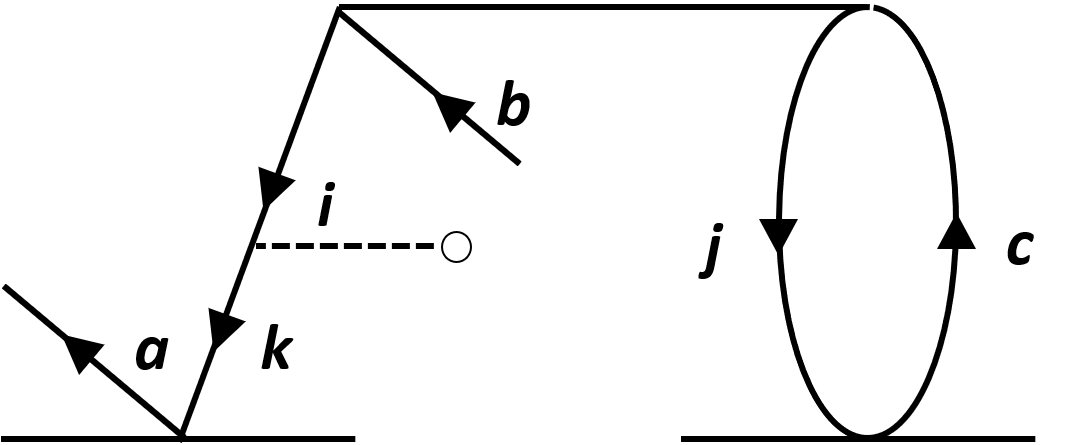}\\
(xix) & (xx) & (xxi) & (xxii) & (xxiii) & (xxiv)\\ \\
\includegraphics[width=2.2cm,height=1.4cm]{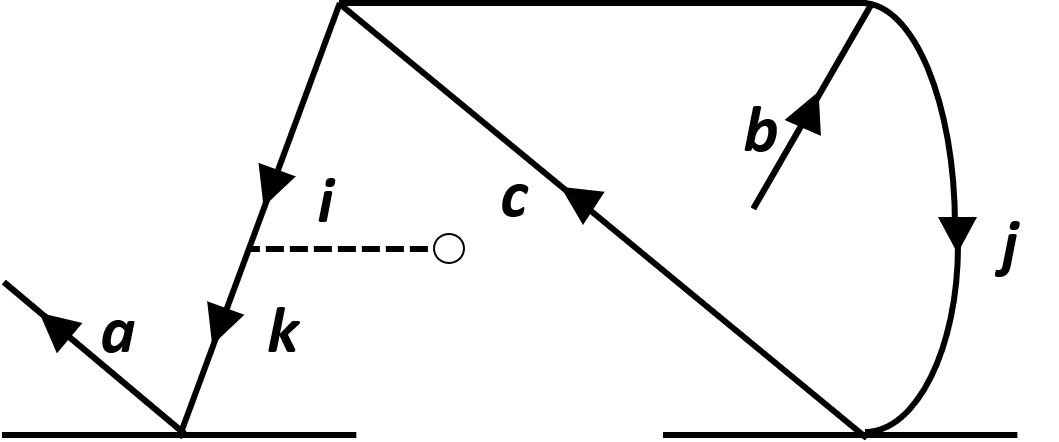} & \includegraphics[width=2.2cm,height=1.4cm]{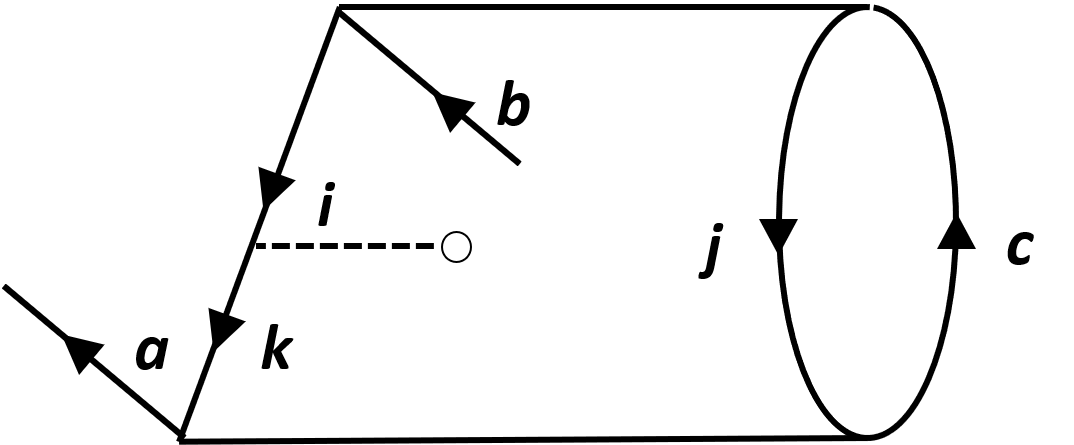} & \includegraphics[width=2.2cm,height=1.4cm]{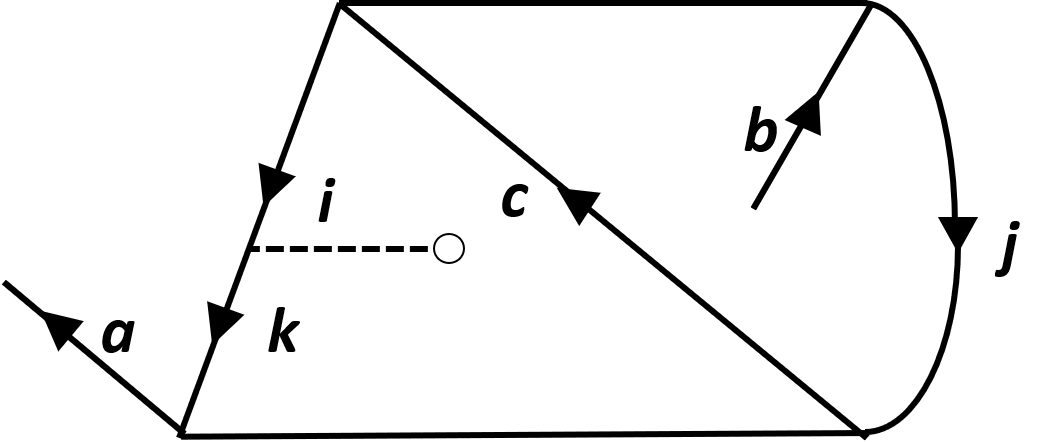} &
&&\\
(xxv) & (xxvi) & (xxvii) & & & \\ 
\end{tabular}
\caption{The effective one-body terms representing particle-particle (p-p) diagrams considered in this work. $i, j, k, \cdots$ and $a, b, c, \cdots$ refer to holes and particles, respectively. The symbol of the operator, $O_{p-p}$, is not mentioned explicitly in the diagrams, and the property vertex is the dashed line ending with an `$o$' in each diagram. }
\label{fig:figure2}
\end{figure*}

\begin{figure*}[t]
\centering
\begin{tabular}{cccccc}
\includegraphics[width=1.4cm,height=1.4cm]{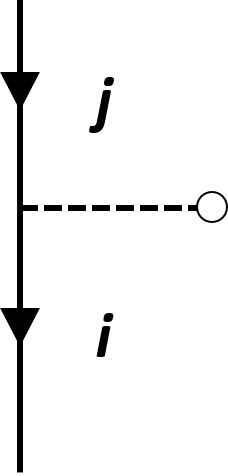} & \includegraphics[width=1.6cm,height=1.4cm]{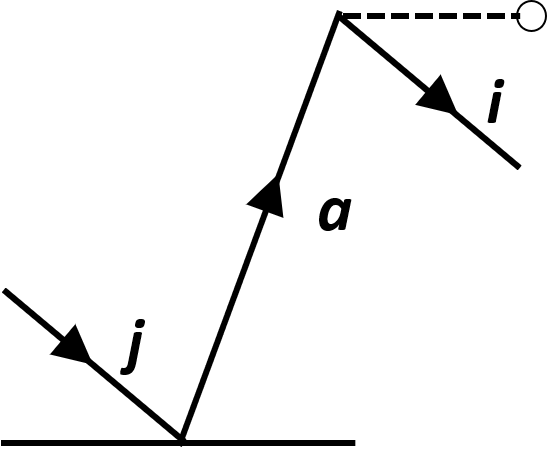} & \includegraphics[width=1.6cm,height=1.4cm]{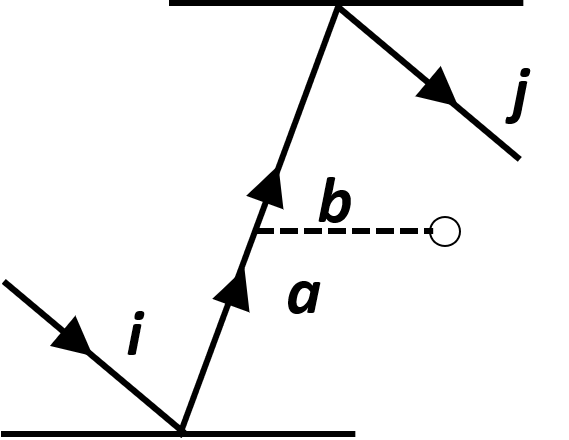} &
\includegraphics[width=1.6cm,height=1.4cm]{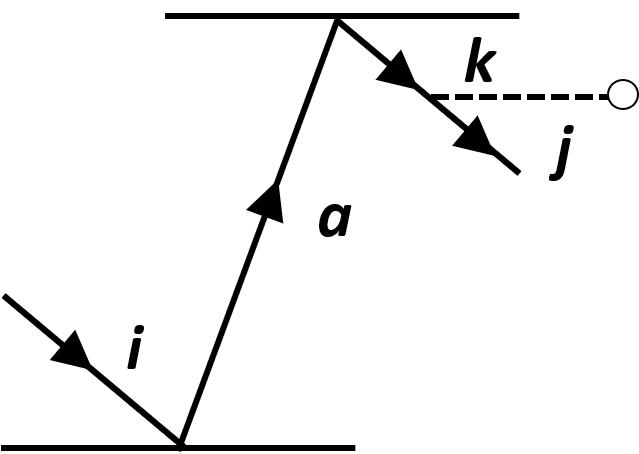} &
\includegraphics[width=1.6cm,height=1.4cm]{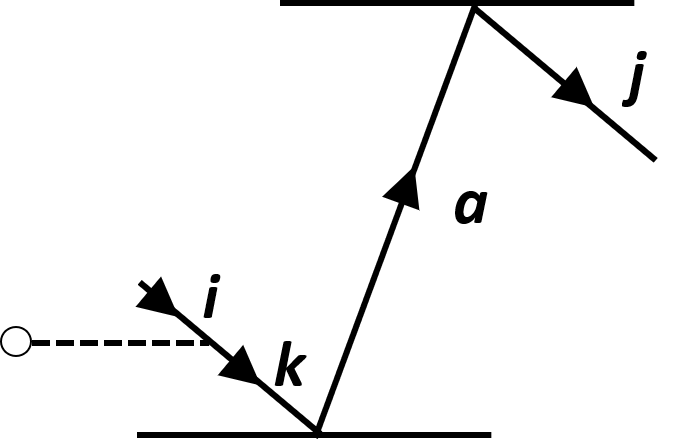} &
\includegraphics[width=2.2cm,height=1.4cm]{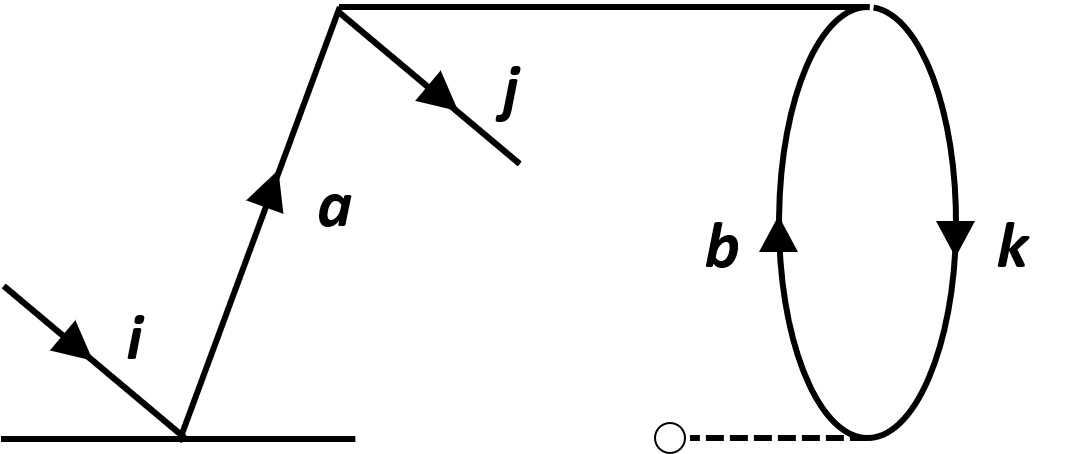}\\
(i) & (ii) & (iii) & (iv) & (v) & (vi) \\ \\
\includegraphics[width=2.2cm,height=1.4cm]{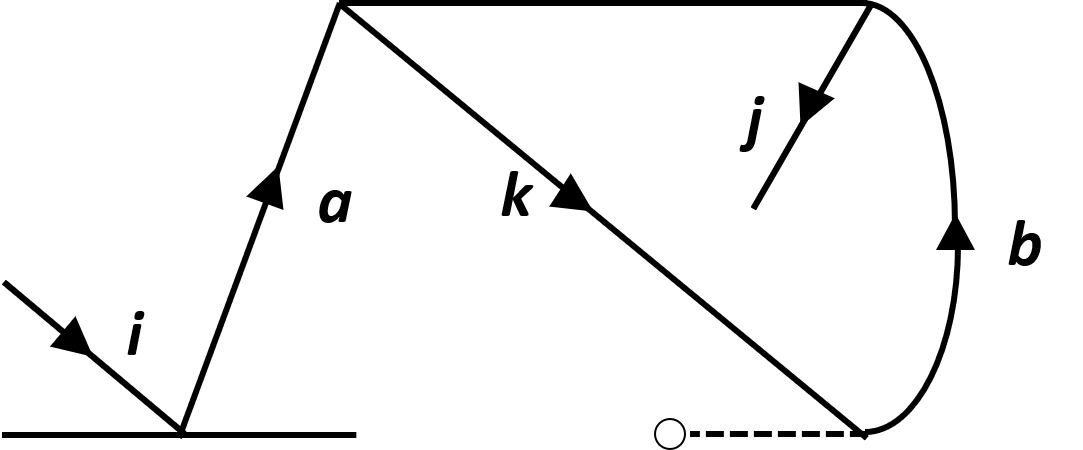} & \includegraphics[width=2.2cm,height=1.4cm]{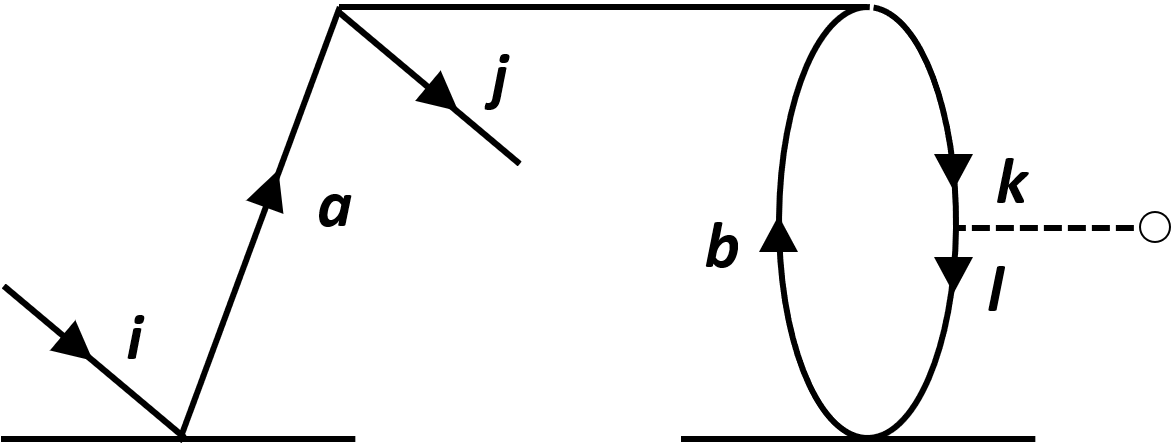} & \includegraphics[width=2.2cm,height=1.4cm]{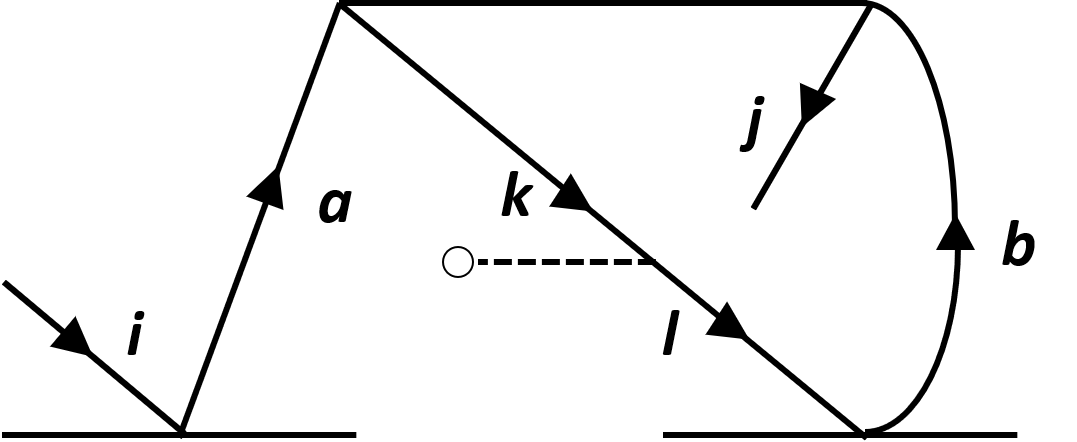} &
\includegraphics[width=2.2cm,height=1.4cm]{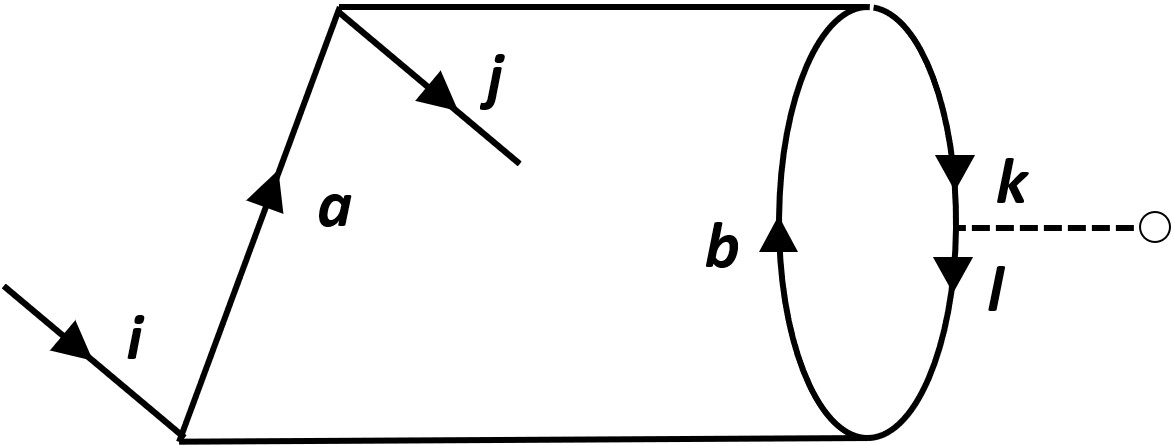} &
\includegraphics[width=2.2cm,height=1.4cm]{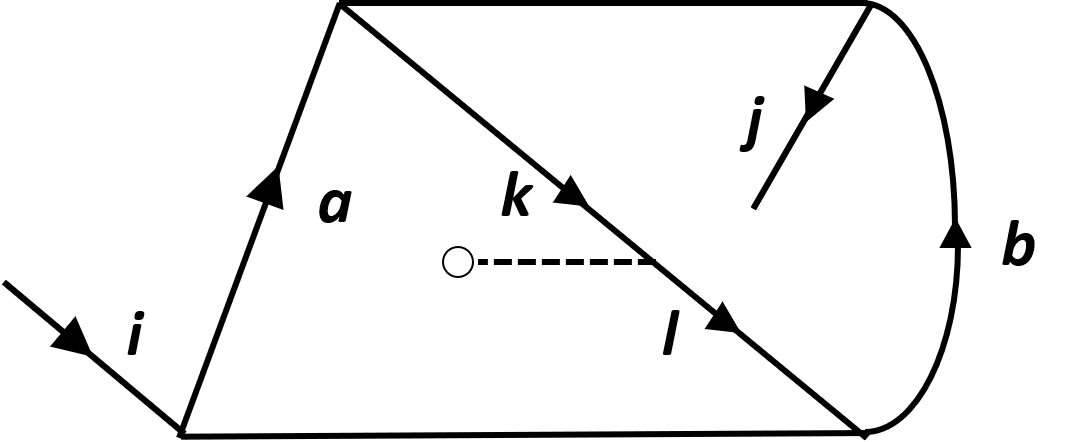} &
\includegraphics[width=2.2cm,height=1.4cm]{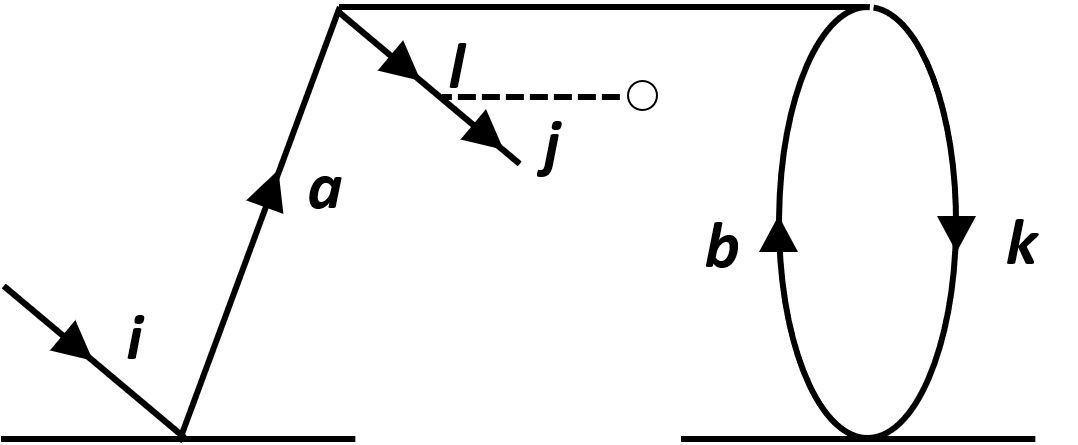}\\
(vii) & (viii) & (ix) & (x) & (xi) & (xii)\\ \\
\includegraphics[width=2.2cm,height=1.4cm]{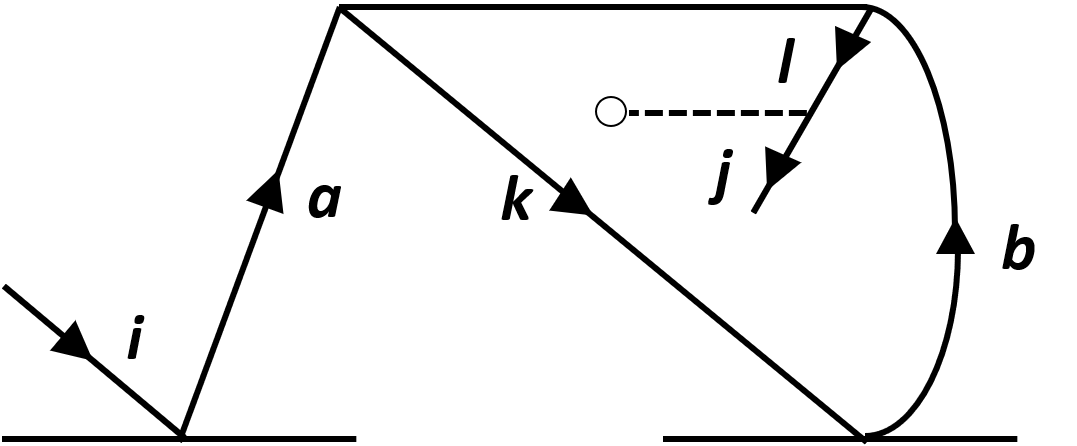} & \includegraphics[width=2.2cm,height=1.4cm]{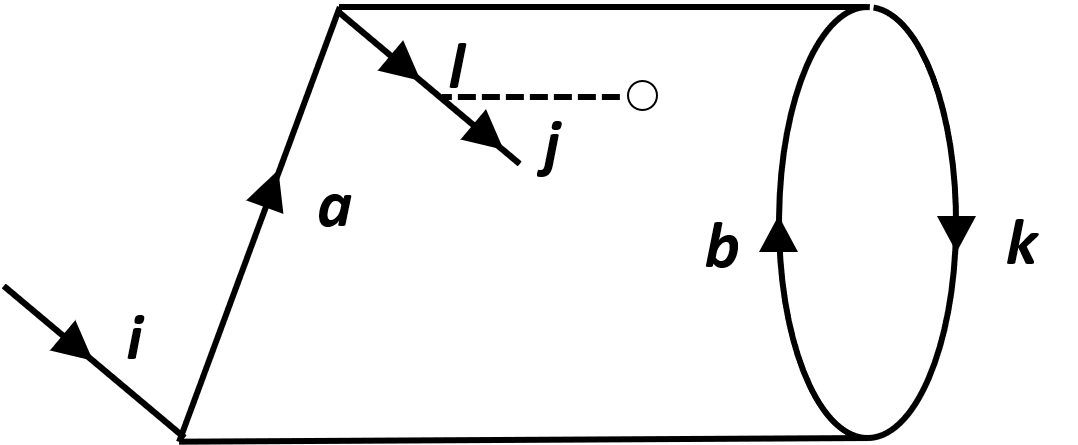} & \includegraphics[width=2.2cm,height=1.4cm]{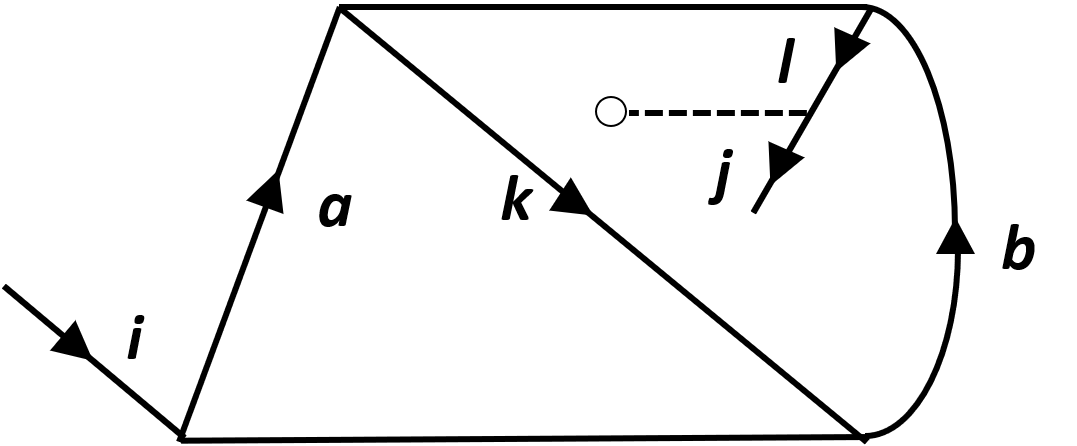} &
\includegraphics[width=2.2cm,height=1.4cm]{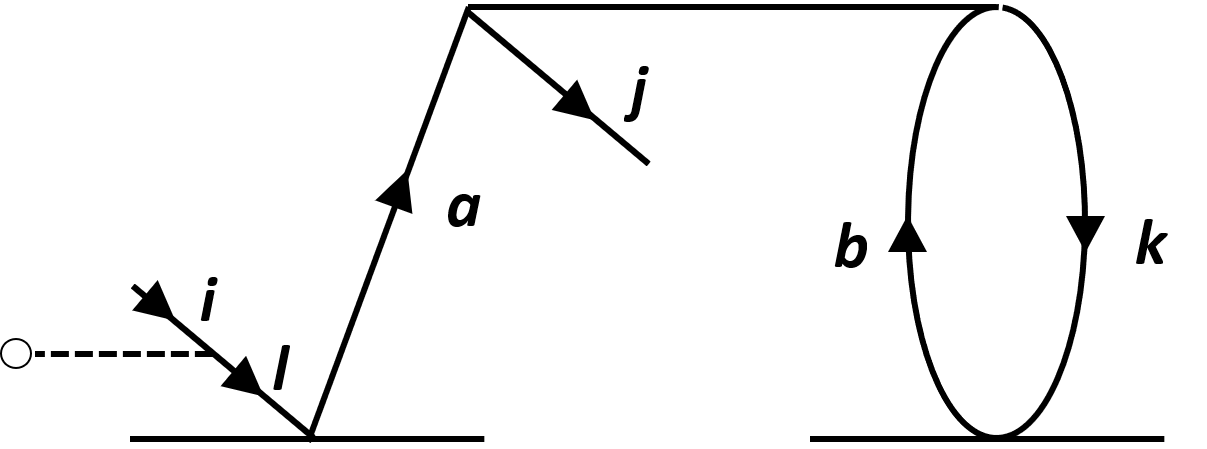} &
\includegraphics[width=2.2cm,height=1.4cm]{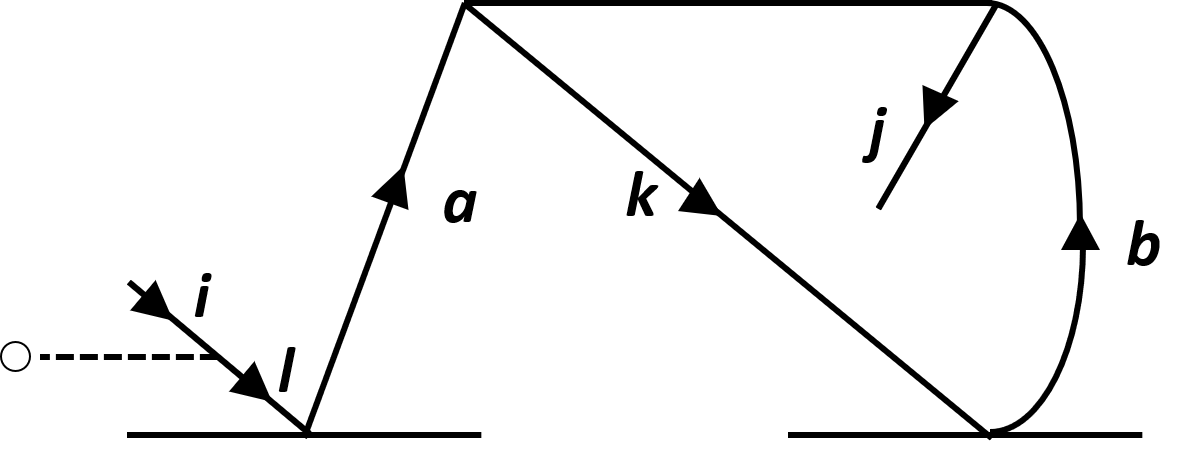} &
\includegraphics[width=2.2cm,height=1.4cm]{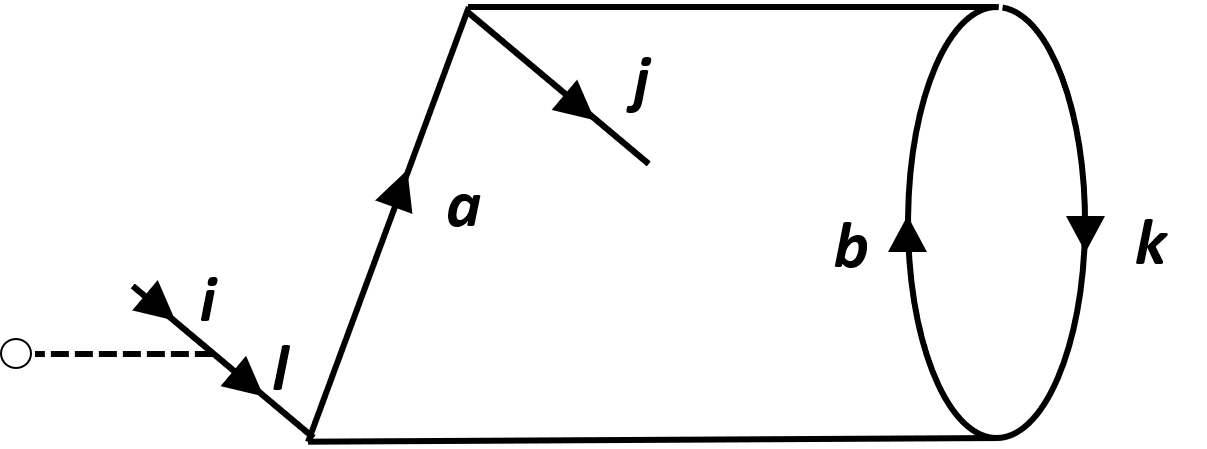}\\
(xiii) & (xiv) & (xv) & (xvi) & (xvii) & (xviii)\\ \\
\includegraphics[width=2.2cm,height=1.4cm]{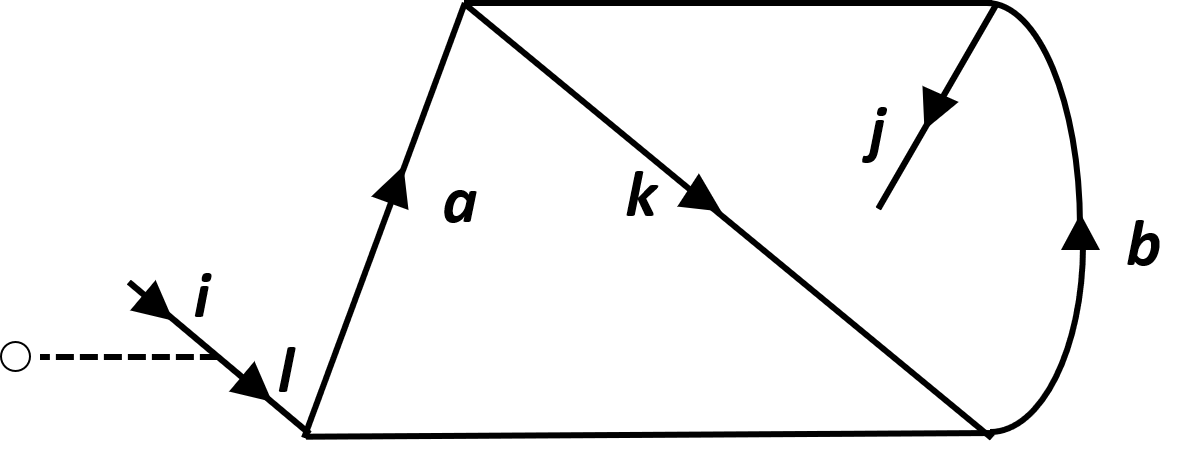} & \includegraphics[width=2.2cm,height=1.4cm]{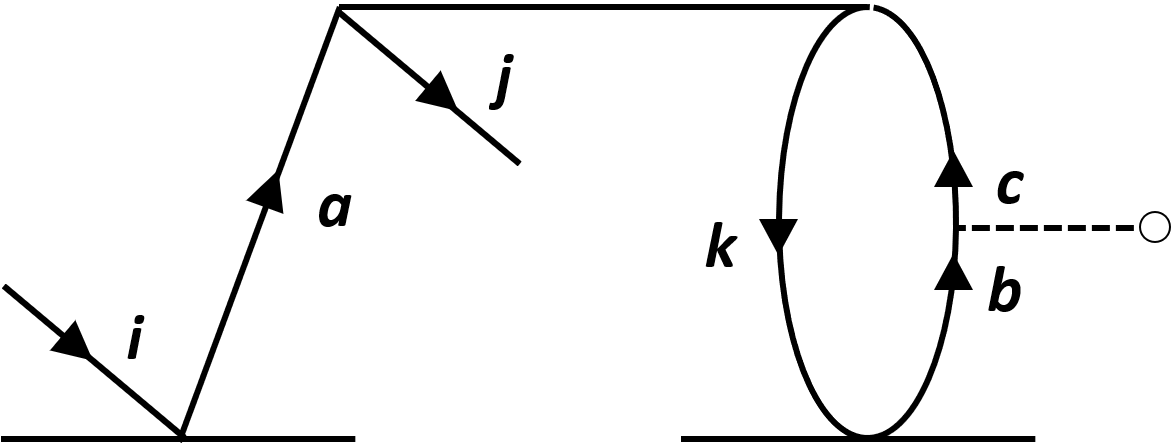} & \includegraphics[width=2.2cm,height=1.4cm]{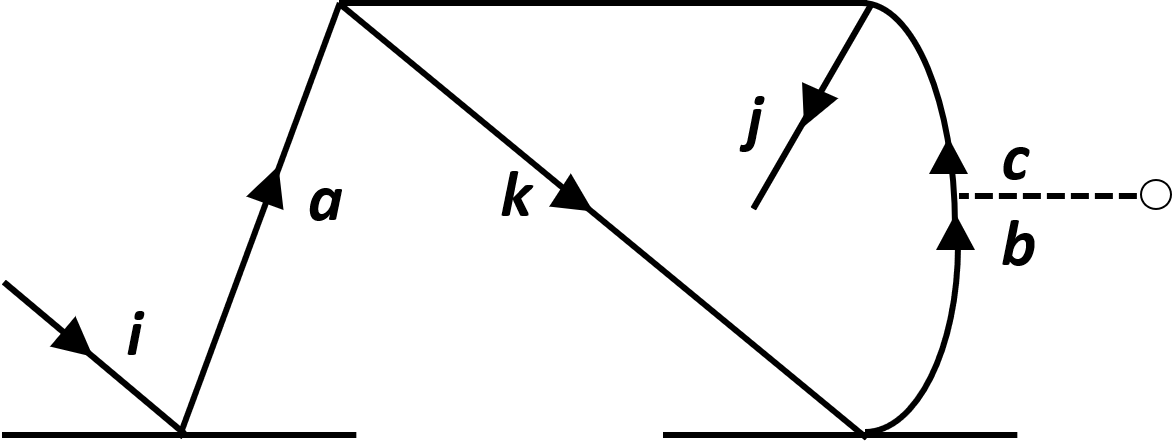} &
\includegraphics[width=2.2cm,height=1.4cm]{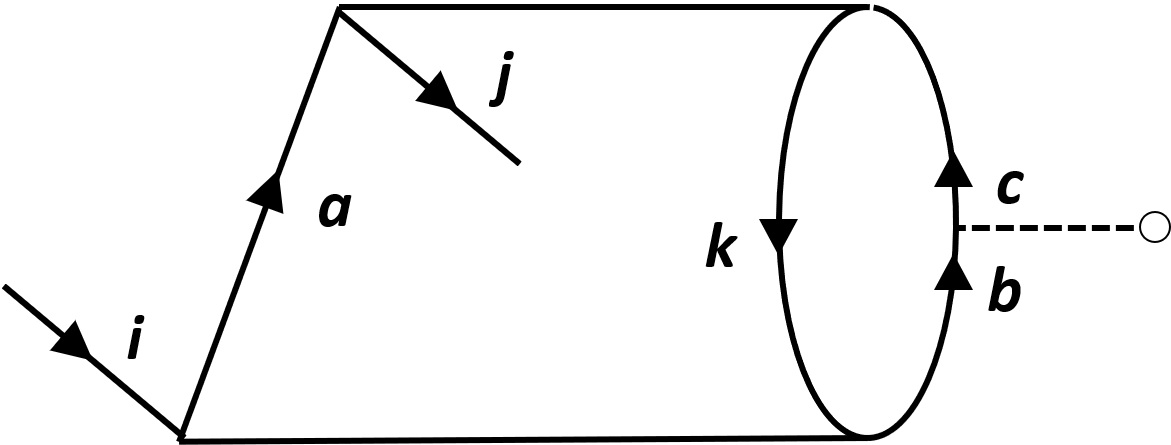} &
\includegraphics[width=2.2cm,height=1.4cm]{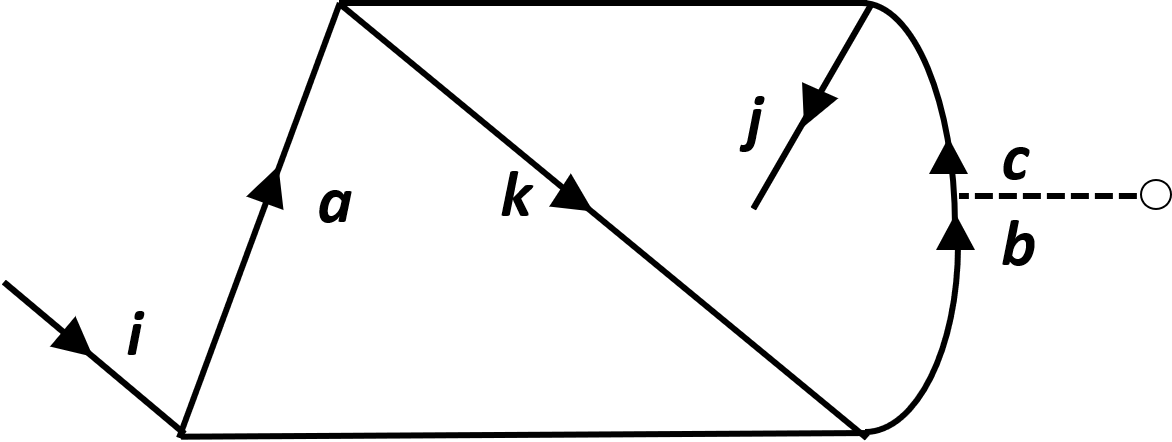} &
\includegraphics[width=2.2cm,height=1.4cm]{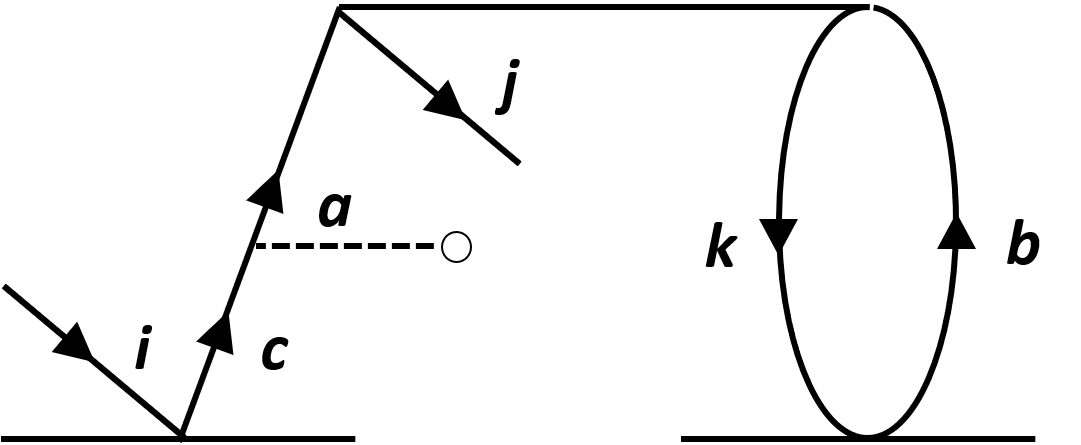}\\
(xix) & (xx) & (xxi) & (xxii) & (xxiii) & (xxiv)\\ \\
\includegraphics[width=2.2cm,height=1.4cm]{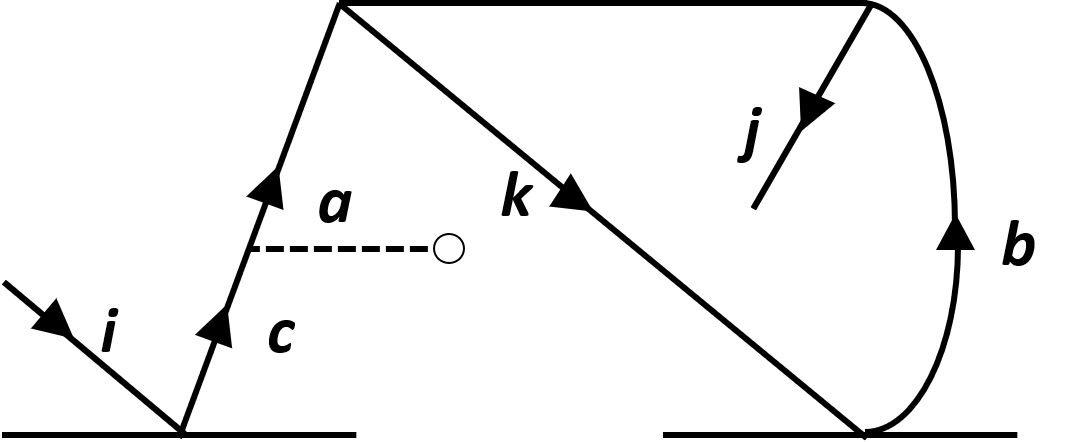} & \includegraphics[width=2.2cm,height=1.4cm]{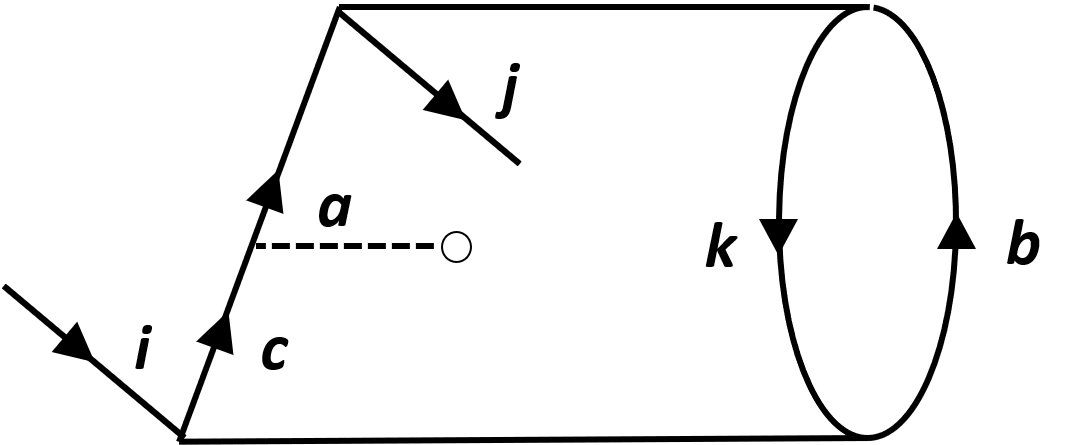} & \includegraphics[width=2.2cm,height=1.4cm]{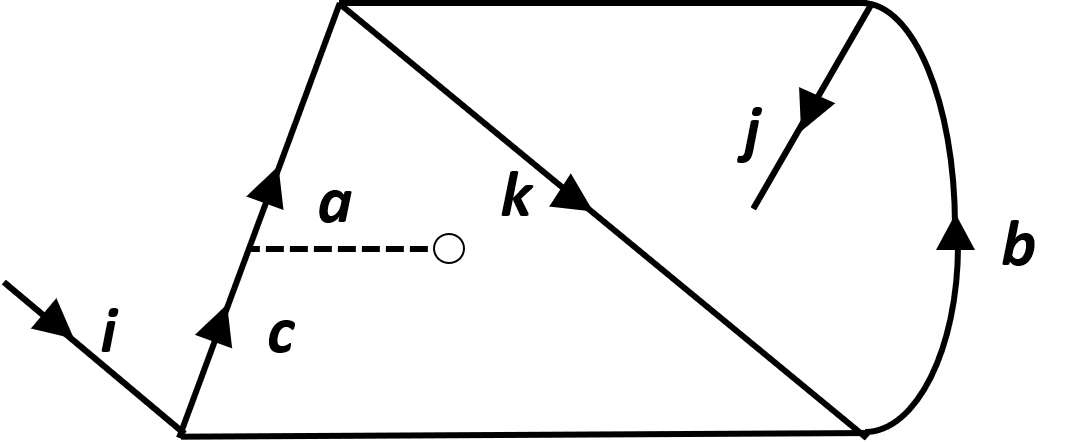} &&&\\
(xxv) & (xxvi) & (xxvii) & & & \\ 
\end{tabular}
\caption{The effective one-body terms representing the hole-hole (h-h) diagrams that are included in this work. The notations are the same as in the figure for the particle-particle diagrams. The property operator, $O_{h-h}$ is not explicitly mentioned in each of the diagrams, just as in Fig. \ref{fig:figure2}. }
\label{fig:figure3}
\end{figure*}

\begin{figure*}[t]
\centering
\begin{tabular}{cccccc}
\includegraphics[width=1.4cm,height=1.4cm]{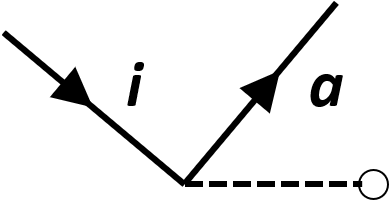} & \includegraphics[width=1.6cm,height=1.4cm]{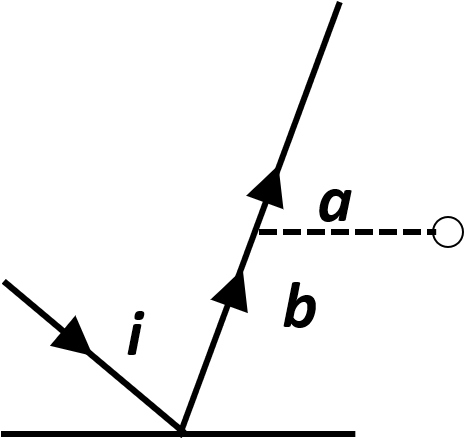} & \includegraphics[width=1.6cm,height=1.4cm]{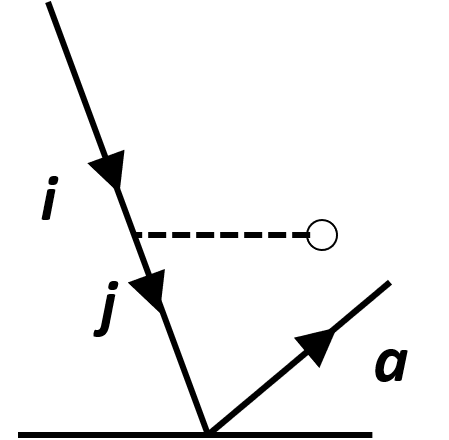} &
\includegraphics[width=2.2cm,height=1.4cm]{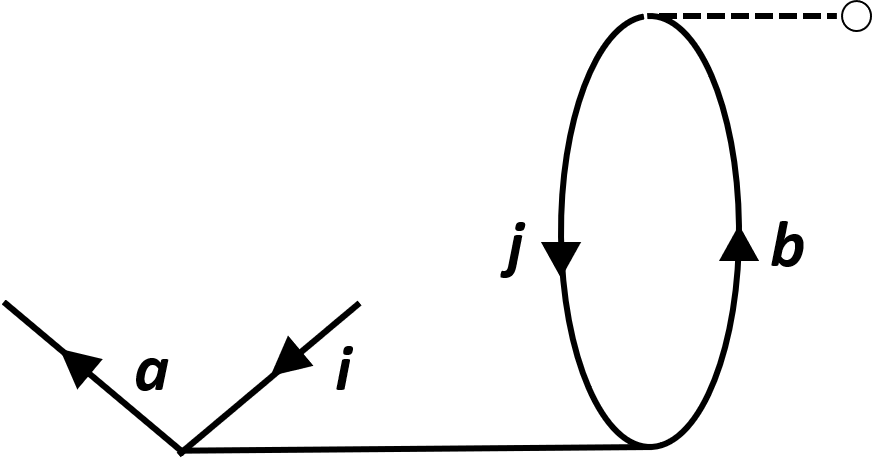} &
\includegraphics[width=2.2cm,height=1.4cm]{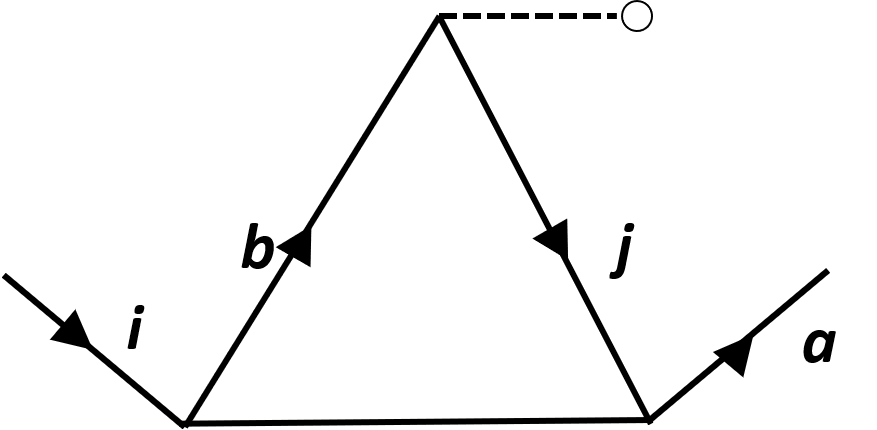} &
\includegraphics[width=2.2cm,height=1.4cm]{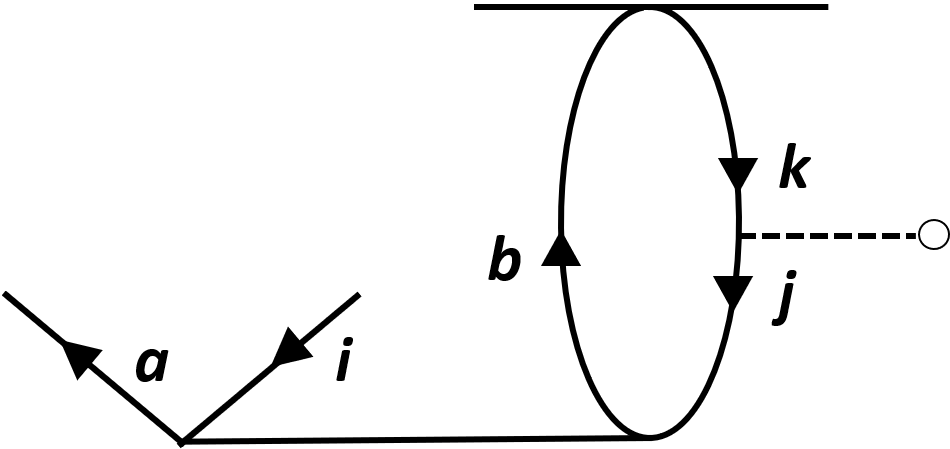}\\
(i) & (ii) & (iii) & (iv) & (v) & (vi) \\ \\
\includegraphics[width=2.2cm,height=1.4cm]{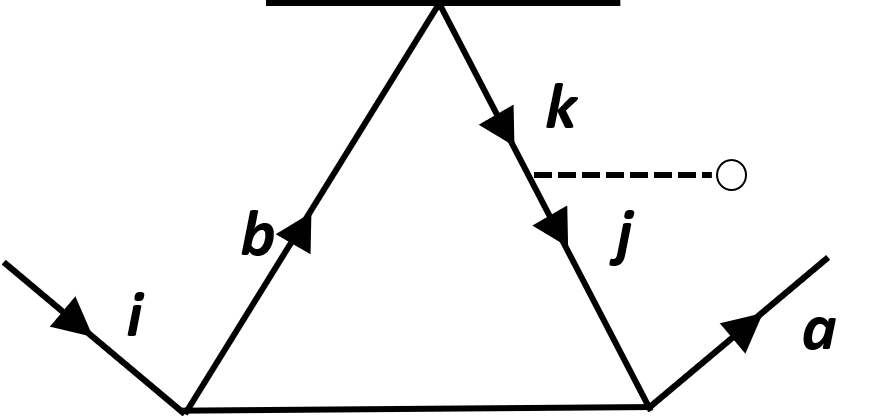} & \includegraphics[width=2.2cm,height=1.4cm]{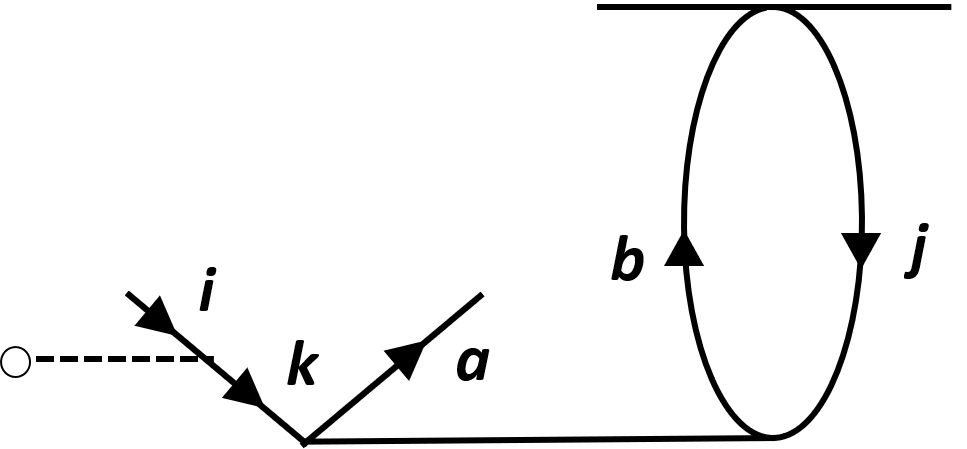} & \includegraphics[width=2.2cm,height=1.4cm]{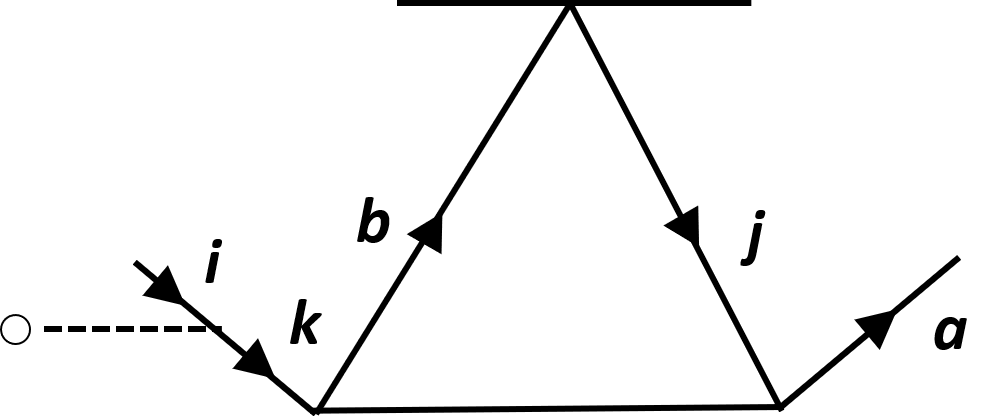} &
\includegraphics[width=2.2cm,height=1.4cm]{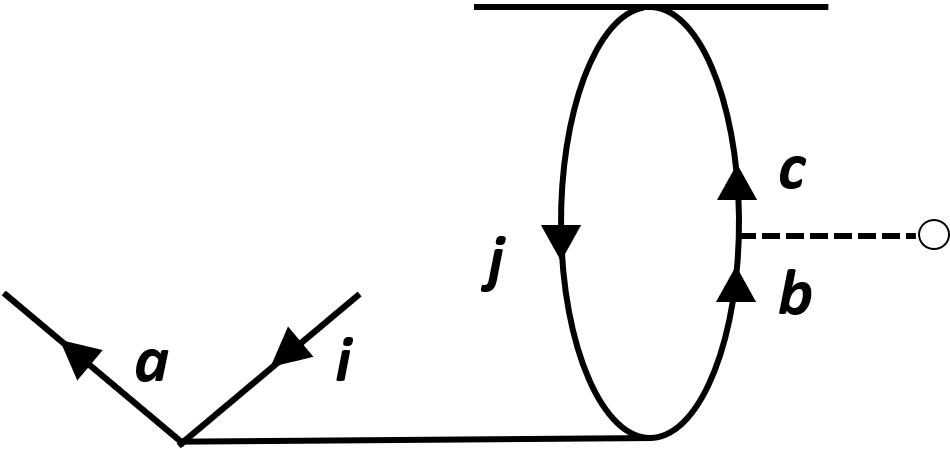} &
\includegraphics[width=2.2cm,height=1.4cm]{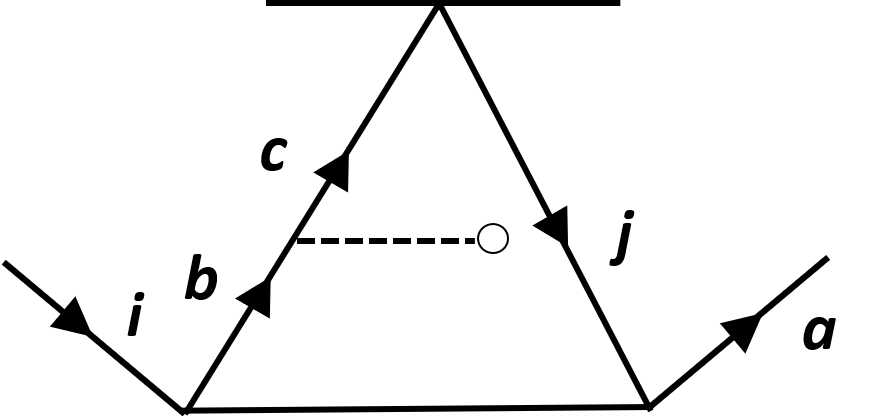} &
\includegraphics[width=2.2cm,height=1.4cm]{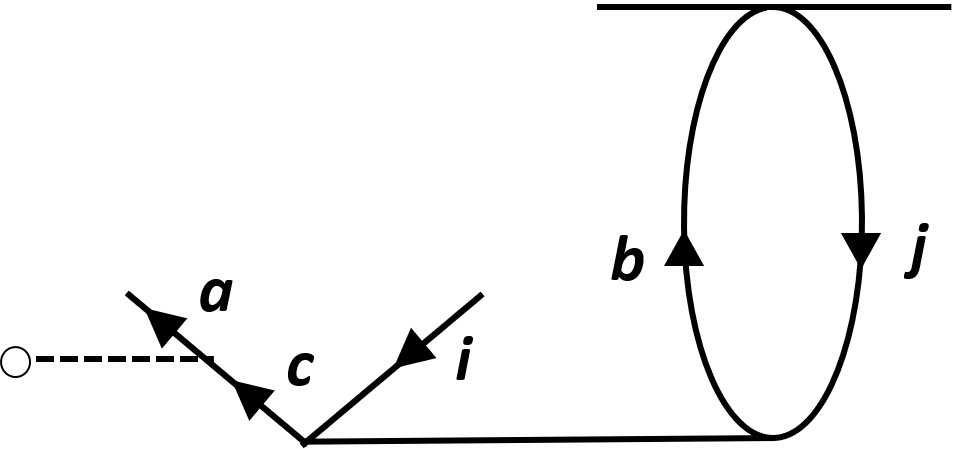}\\
(vii) & (viii) & (ix) & (x) & (xi) & (xii)\\ \\
\includegraphics[width=2.2cm,height=1.4cm]{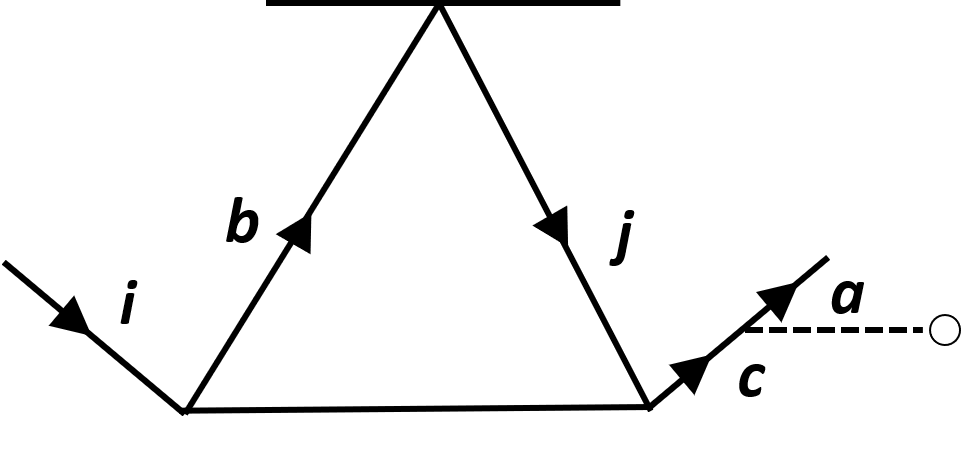} & \includegraphics[width=2.2cm,height=1.4cm]{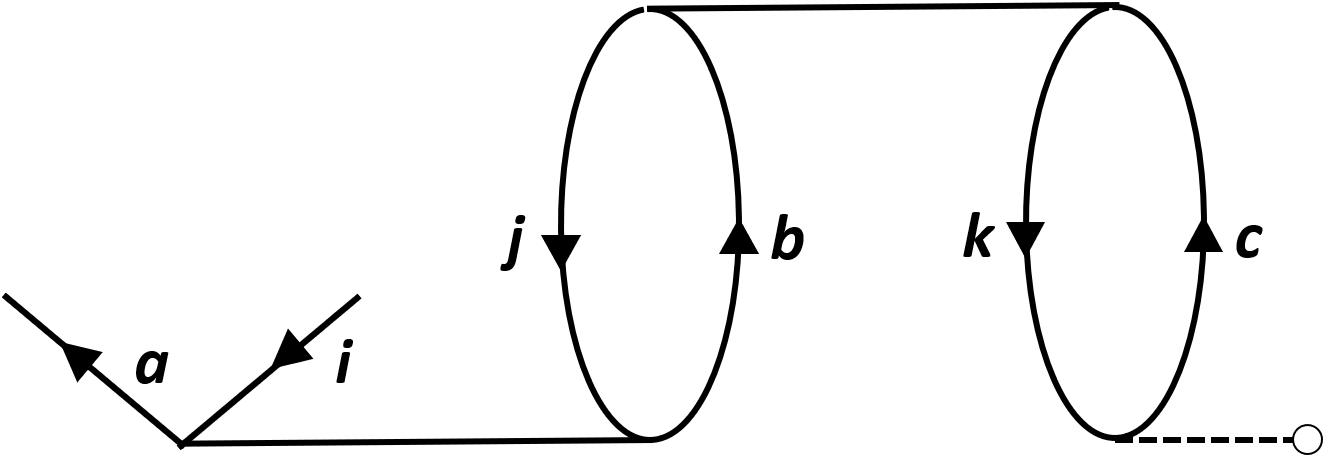} & \includegraphics[width=2.2cm,height=1.4cm]{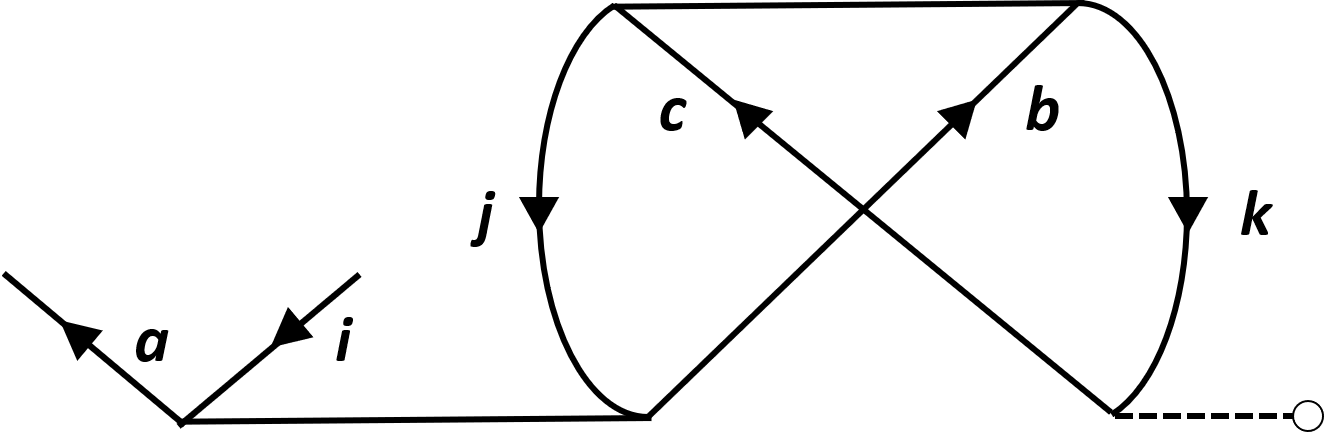} &
\includegraphics[width=2.2cm,height=1.4cm]{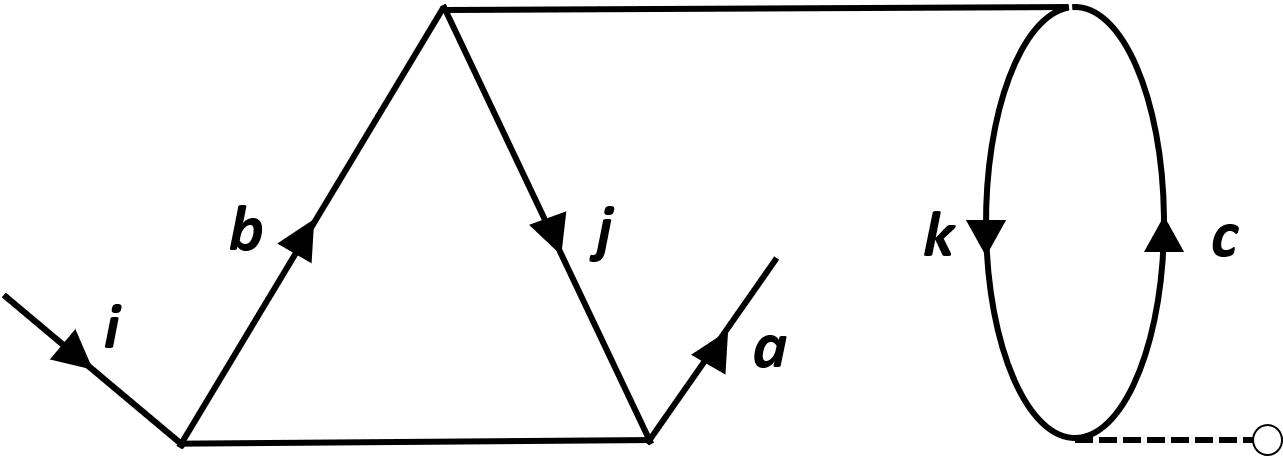} &
\includegraphics[width=2.2cm,height=1.4cm]{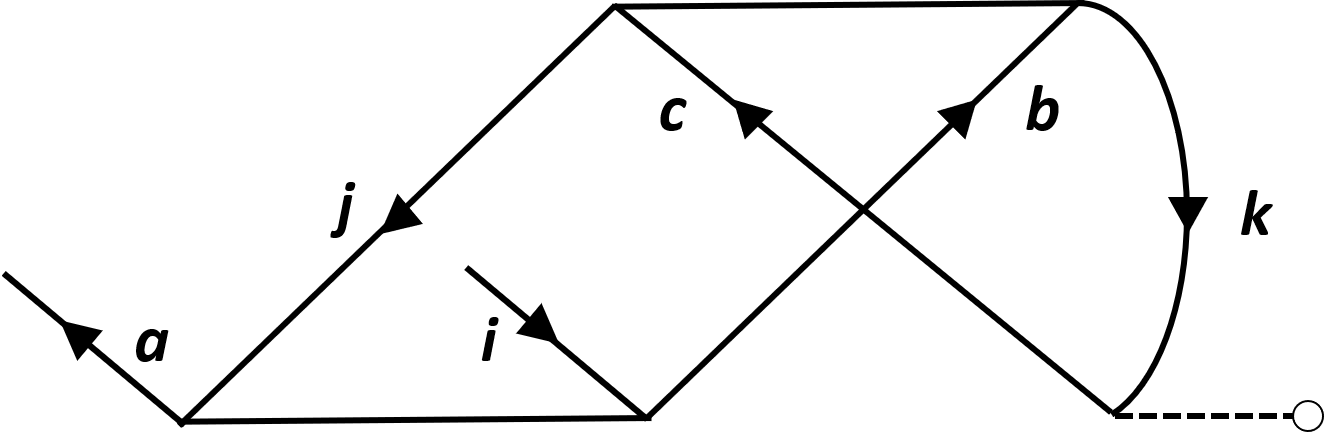} &
\includegraphics[width=2.2cm,height=1.4cm]{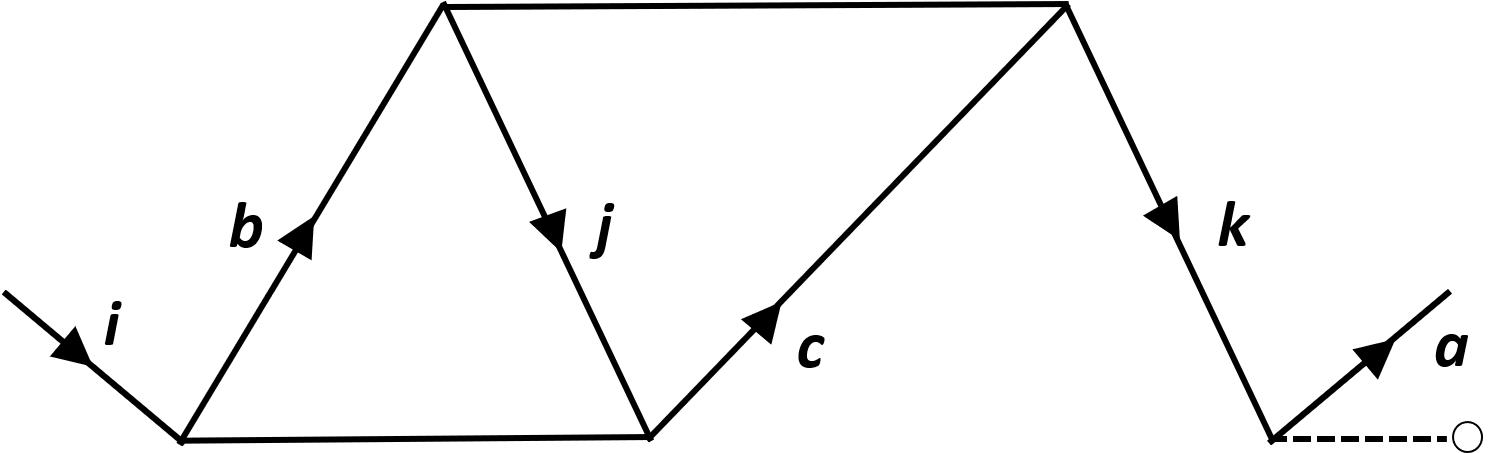}\\
(xiii) & (xiv) & (xv) & (xvi) & (xvii) & (xviii)\\ \\
\includegraphics[width=2.2cm,height=1.4cm]{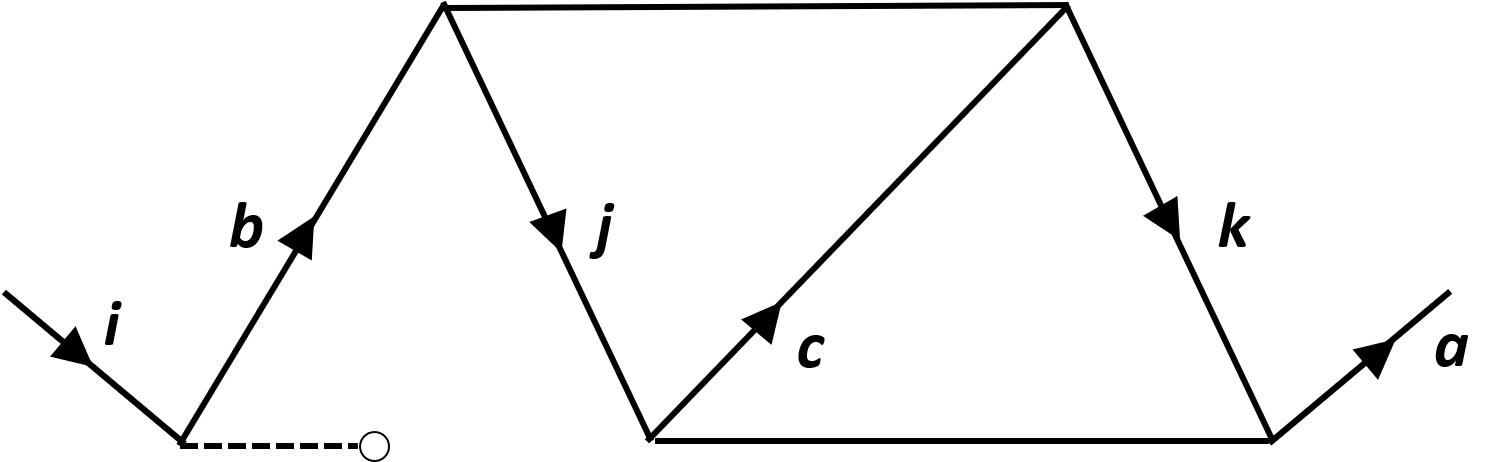} & \includegraphics[width=2.2cm,height=1.4cm]{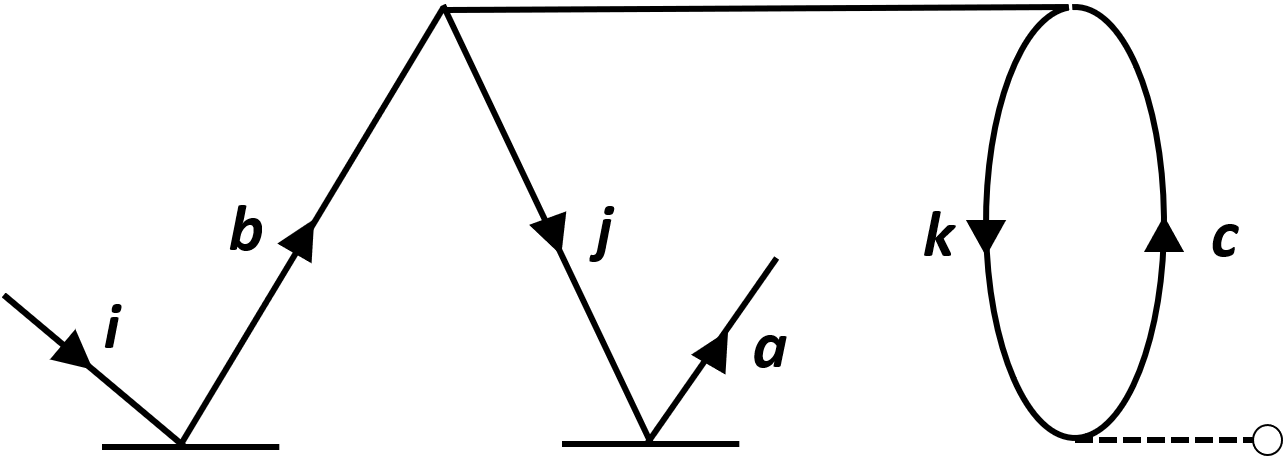} & \includegraphics[width=2.2cm,height=1.4cm]{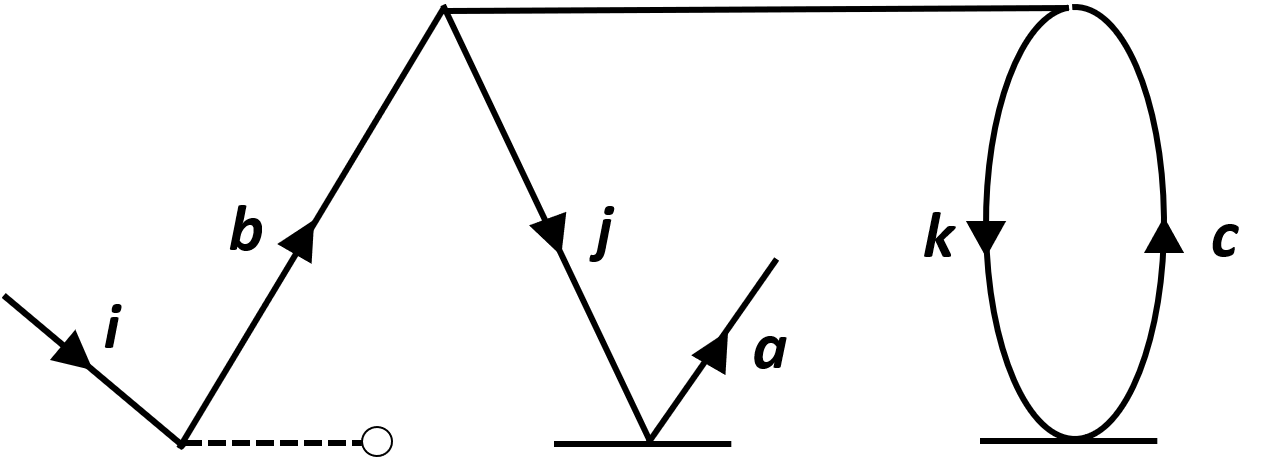} &
\includegraphics[width=2.2cm,height=1.4cm]{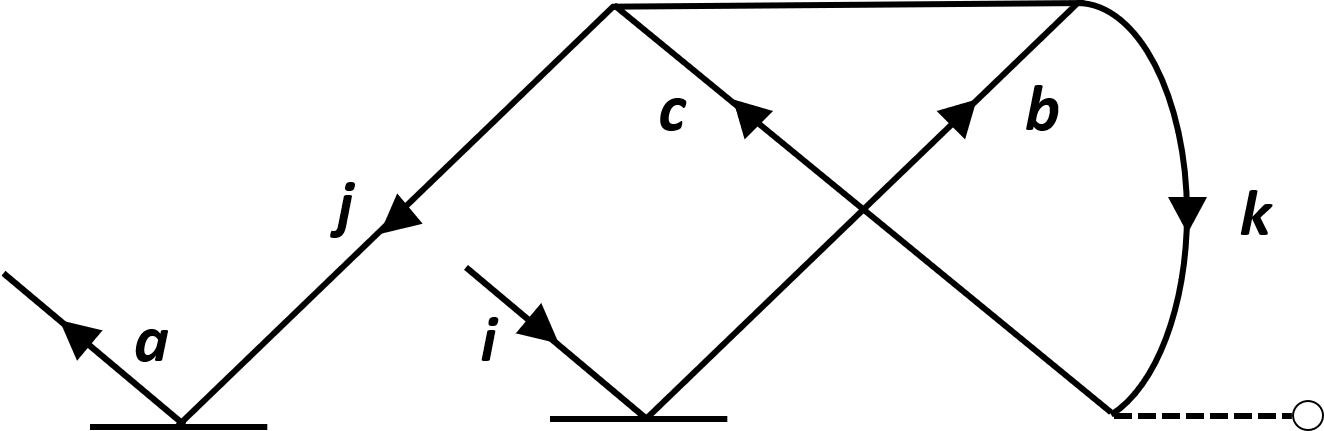} &
\includegraphics[width=2.2cm,height=1.4cm]{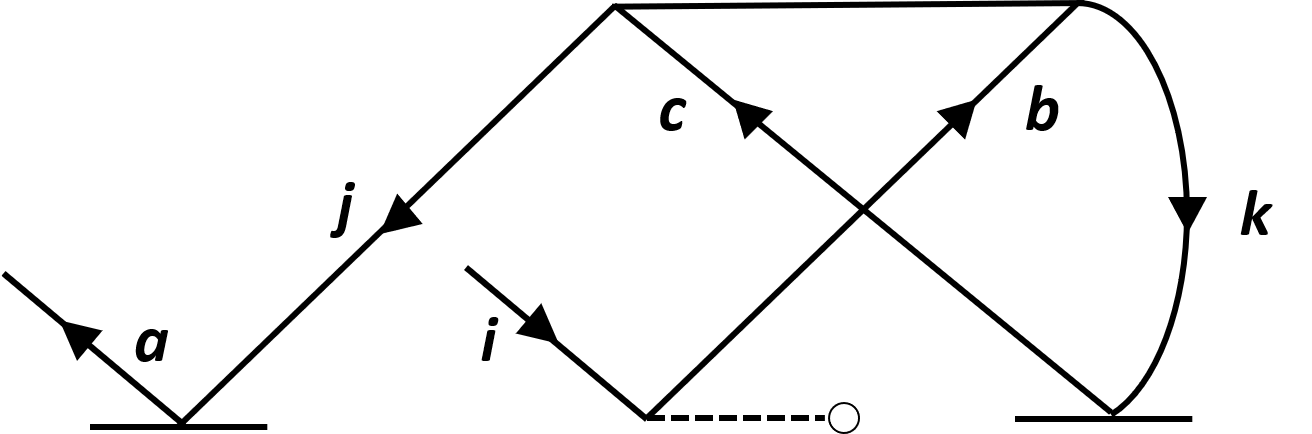} &
\includegraphics[width=2.2cm,height=1.4cm]{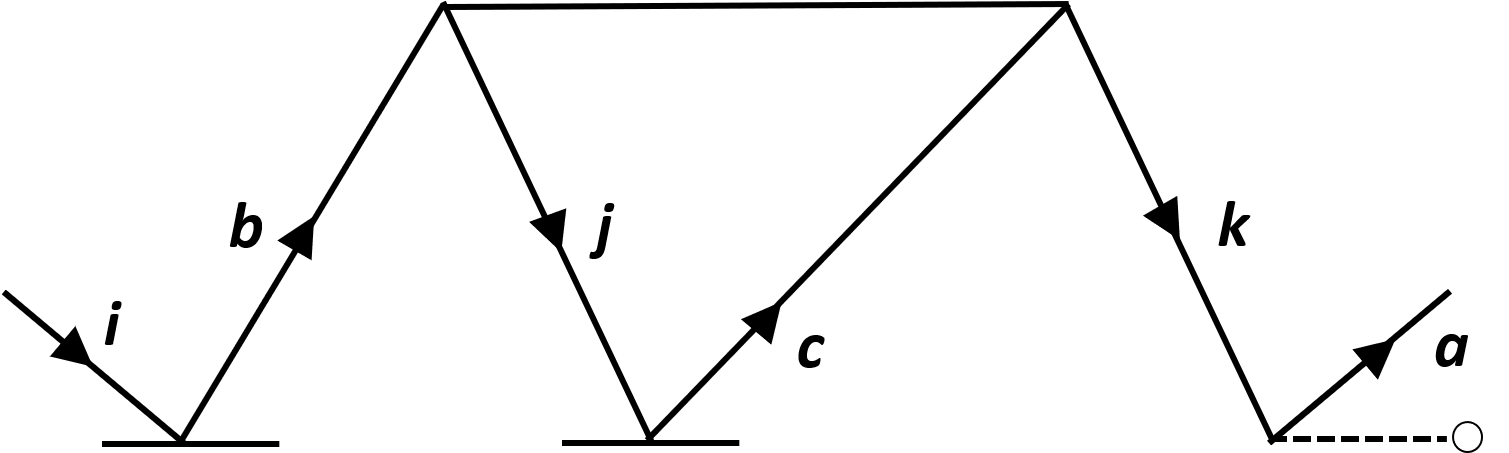}\\
(xix) & (xx) & (xxi) & (xxii) & (xxiii) & (xxiv)\\ \\
\includegraphics[width=2.2cm,height=1.4cm]{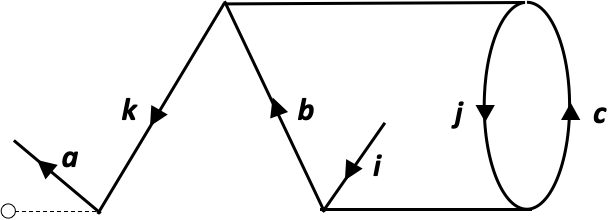} & \includegraphics[width=2.2cm,height=1.4cm]{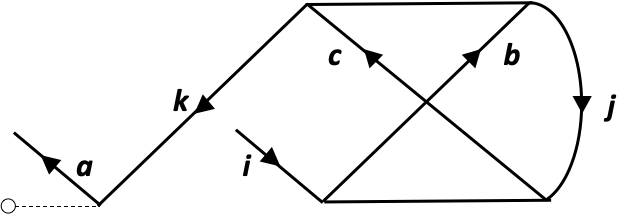} &
\includegraphics[width=2.2cm,height=1.4cm]{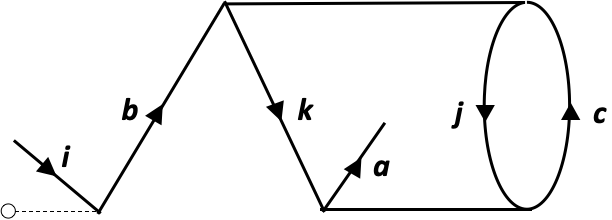}&
\includegraphics[width=2.2cm,height=1.4cm]{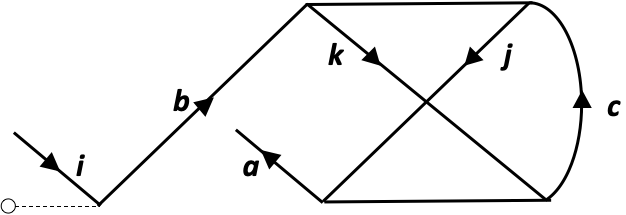}&
&\\
(xxv) & (xxvi) & (xxvii)&(xxviii) & & \\ 
\end{tabular}
\caption{The list of the effective one-body terms representing the particle-hole (p-h) diagrams in this work. The notations are the same as in the particle-particle and the hole-hole diagrams. }
\label{fig:figure4}
\end{figure*}

The expectation value of an operator, $O$, in the (R)CC method, can be written as follows 
\begin{eqnarray}
 \langle   {O} \rangle = \frac{\langle \Psi \arrowvert O \arrowvert \Psi \rangle}{\langle \Psi \arrowvert \Psi \rangle} 
 &=& \frac{\langle \Phi_0 \arrowvert e^{T^\dag} {O} e^T \arrowvert \Phi_0 \rangle}{\langle \Phi_0 \arrowvert e^{  T^\dag} e^T \arrowvert \Phi_0 \rangle} \nonumber \\
 &=& \langle \Phi_0 \arrowvert (e^{T\dag} O e^T) \arrowvert \Phi_0 \rangle_c, 
 \label{eqpr}
\end{eqnarray}
where  the subscript, `$c$', means that each term is fully contracted~\cite{Kvas}, or in the diagrammatic terminology, connected \cite{Bartlett}. 

The PDM of a molecule is determined as \cite{AEM} 
\begin{eqnarray}
 \mu = \langle \Psi \arrowvert D \arrowvert \Psi \rangle + \sum_A Z_A r_A, 
\end{eqnarray}
where $D$ is the electric dipole operator, the index $A$ runs over the number of nuclei, $Z_A$ is the atomic number of the $A^{th}$ nucleus, and $r_A$ is the position vector from the origin to the site of the $A^{th}$ nucleus. The first term in the above expression is the electronic term, while the second term is the nuclear contribution. 

Similarly, $\mathcal{E}_{\mathrm{eff}}$ is evaluated as 
\begin{eqnarray}
 \mathcal{E}_{\mathrm{eff}} 
 &=& \langle \Psi \arrowvert \sum_{i=1}^{N_e} \beta  \Sigma_i \cdot  E^{\mathrm{intl}}_i \arrowvert \Psi \rangle ,
\end{eqnarray}
where the summation is over the number of electrons, $N_e$, $\beta$ is one of the Dirac matrices (also known as $\gamma_0$ in literature), $\Sigma$ is the (4x4) version of Pauli matrices, and  $E^{\mathrm{intl}}_i$ is the internal electric field that is experienced by the $i^{th}$ electron, and is given by the negative of the gradient of the sum of electron-nucleus and electron-electron interaction potentials. Since the expression given above involves evaluating integrals over a two-body Coulomb operator, $\frac{1}{r_{ij}}$, and is complicated, we take recourse to employing an effective eEDM Hamiltonian instead of the one introduced above~\cite{BPD}. It follows that 
\begin{eqnarray}
\mathcal{E}_{\mathrm{eff}} = - 2ic  \langle \Psi \arrowvert \sum_{i=1}^{N_e} \beta \gamma_5 p_i^2 \arrowvert \Psi \rangle ,
\end{eqnarray}
where $\gamma_5$ is the product of the gamma matrices (given by $i\gamma_0\gamma_1\gamma_2\gamma_3$), while $p_i$ is the momentum of the $i^{th}$ electron. 

In the LERCCSD method, the following expression has been used in the evaluation of the expectation values
\begin{eqnarray}
 \langle   {O} \rangle = \langle \Phi_0 \arrowvert (1+T_1+T_2)^{\dag} O (1+T_1+T_2) \arrowvert \Phi_0 \rangle_c.
\end{eqnarray}

These terms are represented using Goldstone diagrams and are shown in Fig. \ref{fig:figure1}. Note that $OT_2$ and its hermitian conjugate are zero, due to Slater-Condon rules \cite{Slater,Condon}.  Diagrammatically, such a diagram will have at least two open lines, that is, it is not fully  connected. The evaluation of properties using the LERCCSD approximation misses contributions corresponding to many correlation effects that will arise from the relativistic third-order many-body perturbation theory (RMBPT method). On the other hand, it is not possible to evaluate exactly Eq. (\ref{eqpr}) even in the RCCSD method as it contains a non-terminating expression. However, it is possible to demonstrate the importance of contributions from the leading-order non-linear terms corresponding to the third- and fourth-order effects of the RMBPT method. It is still extremely challenging to perform direct calculations by incorporating the non-linear terms of Eq. (\ref{eqpr}) in heavier molecules, due to amount of computations involved in it. In order to tackle this issue, we adopt an additional computational step by breaking the non-linear terms into intermediate parts as described more elaborately below. Further, we parallelize the program using Message Passing Interface (MPI) and show the scalability of their calculations with the number of processors of a computer. 

The approach can be understood by revisiting the diagrams in Fig. \ref{fig:figure1}. As an example, we consider sub-figure (ii). The property vertex has one incoming particle line and an outgoing hole line. We define it as a particle-hole vertex. Such particle-hole vertices can be found in sub-figures (v) to (viii) too. In the intermediate-diagram formalism, the vertex $O$ is removed and replaced successively by each of the particle-hole (p-h) diagrams (more precisely, their hermitian conjugates) of Fig. \ref{fig:figure4}. We assign the notation $O_{p-h}$ for such a vertex. This sequence of operations already generates 26 diagrams from $O_{p-h}T_1$, and includes terms that occur in the RMBPT method. We note that $O_{p-h}T_1$ in the case of hermitian conjugate of sub-figure (i) of Fig \ref{fig:figure4} gives back the LERCCSD diagram for $OT_1$. Similar to $O_{p-h}$, we also construct the analogous $O_{h-h}$ and $O_{p-p}$ diagrams (as given in Figs. \ref{fig:figure2} and \ref{fig:figure3}, respectively), and generate more terms. We note that the property vertex from the DF diagram in Fig. \ref{fig:figure1} is \textit{not} replaced with any intermediate diagram as otherwise there will be repetition of diagrams in the calculations. Further, one has to be careful to avoid repetition of diagrams while contracting effective $O$ operators with the $T$ RCC operators. For example, it can be shown that $T_1^{\dagger}OT_1$ diagrams can appear twice through $T_1^{\dagger}O_{p-h}$ and $T_1^{\dagger}O_{p-p/h-h}T_1$ terms. Such diagrams are identified by careful analysis and their double counting is removed manually. 

As can be seen from the above discussions, some of the diagrams that are undertaken in this procedure demand up to the order of $n_h^3n_p^3$ in computational cost for $n_h$ number of holes and $n_p$ number of particles. Therefore, the intermediate diagram approach systematically takes into account non-linear terms while simultaneously cutting down drastically on the computational cost as compared to a direct brute-force evaluation of a non-linear expectation value expression. This can be understood by choosing an example as follows. Replacing $O$ of Fig.~\ref{fig:figure1}(v) by the property vertex with sub-figure (xxv) from Fig.~\ref{fig:figure4} entails a computational cost $~\mathcal{O}(n_p^4n_h^4)$ for the direct evaluation of such a diagram. However, the intermediate-diagram approach leads to a cost of $~\mathcal{O}(n_p^2n_h^2+n_p^3n_h^3)$. This becomes especially relevant when we perform computations on heavy systems and with high-quality basis sets, such as those that we have chosen for this work. For instance, the RCC calculations on HgF involved $n_h = 89$, and $n_p=429$, and therefore the computational cost with an intermediate-diagram approach is a full 5 orders of magnitude smaller than a brute-force approach to computing the same diagram (without considering any molecular point group symmetries). A similar level of reduction in computational cost can be seen from the heaviest HgI too. We add at this point that we have exploited the $C_8$ double point group symmetry in our nLERCCSD code, as we had done for the earlier LERCCSD program~\cite{MA,Yanai}. This aspect also substantially lessens the computational cost, as it restricts the number of matrix elements to be evaluated based on group theoretic considerations. For example, $OT_1$ involves computing matrix elements of the form $\langle a \arrowvert O \arrowvert i \rangle$. Given that we have 89 holes and 429 particles, the number of possible matrix elements are $\sim 3.8 \times 10^5$. However, since we impose the restriction that both $i$ and $a$ should belong to the same irreducible representation, we need to evaluate only $\sim 7.2 \times 10^4$ matrix elements. Similar considerations for the more complicated terms involving $T_2$ leads to evaluating much fewer matrix elements. 

\section{Results and Discussions}\label{sec3}

To carry out the calculations in the considered molecules, we have chosen values for the bond lengths as 2.00686 $A^{\circ}$, 2.42 $A^{\circ}$, 2.62 $A^{\circ}$, 2.81 $A^{\circ}$, 2.075 $A^{\circ}$, and 2.16 $A^{\circ}$ for HgF, HgCl, HgBr, HgI, SrF and BaF, respectively \cite{HgFR,HgXR,SrFR1,SrFR2,BaFR1,BaFR2}. It is to be noted that the chosen values for the HgX molecules are from theory, while those for SrF and BaF are from experiment. Also, we opted for Dyall's quadruple zeta (QZ) basis for Hg and I~\cite{DyallHg}, Dunning's correlation consistent polarized valence quadruple zeta (cc-pVQZ) basis for the halide atoms (F, Cl, and Br)~\cite{Dunning}, and Dyall's QZ functions augmented with Sapporo's diffuse functions~\cite{Sap} for Sr and Ba. We chose Dyall's basis for Hg and I as it is among the most reliable and widely used basis functions for heavy atoms. We did not add diffuse functions as it increases the computational cost drastically for QZ quality basis sets. Moreover, it was found that inclusion of diffuse functions change the effective electric field by around 2.5 percent for HgF, and is expected to lead to a similar difference for the heavier HgX systems~\cite{FFCC}. However, in the foreseeable future, such computations could be peformed to improve the calculated values of the PDMs. To minimize steep computational costs that we incurred due to our choice of QZ basis sets as well as performing all-electron calculations, we cut-off the high-lying virtuals above 1000 atomic units (a.u.) for HgX and BaF. At such a high cut-off value, we can expect that the missing contributions would be minimal, and possibly even negligible. 

In Table~\ref{tab:table1}, we present our results for HgX, SrF, and BaF, all using QZ basis sets. We discuss the trends in the PDMs and  $\mathcal{E}_\mathrm{eff}$s across HgX in the three different approaches, namely LERCCSD, and nLERCCSD methods, while briefly making a comparison with the FFRCCSD method from Ref.~\cite{FFCC} wherever relevant, and also examine the correlation effects in the property from lower to all-order methods. SrF and BaF molecules are treated as stand-alone systems. Firstly, we observe that the effect of non-linear corrections is to increase the PDM and decrease the effective electric field (except in the case of SrF, where the difference is still within 0.5 percent). We find from Table~\ref{tab:table1} that for SrF and BaF, the nLERCCSD method yields PDMs that are very close to their LERCCSD counterparts (within 1.5 percent of each other for both the molecules), but are not in better agreement with experiments than their LERCCSD counterparts. However, the nLERCCSD values agree well with the results from the earlier work that used the FFRCCSD approach (within 1.2 percent of each other) that also employed a QZ quality basis with diffuse functions. Such a comparison cannot be made with the HgX molecules, as available FFRCCSD data uses a double zeta (DZ) quality basis. For HgX systems, we observe that unlike in the cases of SrF and BaF, the difference between the LERCCSD and the nLERCCSD results widen from about 6 percent for HgF and HgCl, to about 25 percent for HgI. The values for $\mathcal{E}_\mathrm{eff}$ for SrF and BaF show that the LERCCSD, nLERCCSD, and FFRCCSD methods all agree to within 1 percent. In the case of HgX molecules, the LERCCSD and the nLERCCSD results are found to differ by at most 2.5 percent. We chose HgF as a representative molecule and performed FFRCCSD calculations with a QZ basis, and found that its effective electric field is 110.87 GV/cm, which is lesser than the nLERCCSD value by 2.5 percent. 

The individual contributions that arise from diagrams given in Fig. \ref{fig:figure1} to the effective electric fields and PDMs of HgX, SrF, and BaF molecules are given in Tables~\ref{tab:table2} and \ref{tab:table3}. The tables give the LERCCSD contributions, where the property vertex is $O$, as well as the nLERCCSD values, where the property vertex could be of the $p-p$, $h-h$, or the $p-h$ type, depending on the diagram. For example, the contribution from sub-figure (ii) of Fig. \ref{fig:figure1}, for the nLERCCSD case involves a p-h vertex, that is, $O_{p-h}T_1$, and therefore includes in it the contributions from the 26 diagrams in Fig. \ref{fig:figure4}. In general, $O$ or $O_{x-y}$ (where `$x$' and `$y$' could be $p$ or $h$) is the eEDM Hamiltonian for $\mathcal{E}_\mathrm{eff}$ (which is given in Table \ref{tab:table2}), while it is the dipole operator for the PDM (which is presented in Table \ref{tab:table3}). 

Table~\ref{tab:table2} shows that for all the systems, the $AT_1$ term always dominates among the correlation terms, where $A$ could correspond to either $OT_1$ or $O_{x-y}T_1$ for LERCCSD or nLERCCSD, respectively. For the effective electric fields of the HgX molecules, in the LERCCSD case, there are strong cancellations among the positive $AT_1$ and the negative $T_1^\dag A T_1$ and $T_2^\dag A T_2$ terms. However, the final values of nLERCCSD and the LERCCSD calculations match within 2.5 percent, since in the former case, the $AT_1$ values are significantly lower than the latter, and the $T_1^\dag A T_1$ sector provides a far smaller contribution. In the case of SrF, the $AT_1$ terms are comparable for LERCCSD and the nLERCCSD scenarios, and therefore inclusion of non-linear terms does not change its effective electric field, while for BaF, we observe a mechanism that is similar to that for the HgX systems. We observe a  different trend for the PDMs, in Table~\ref{tab:table3}. As the DF value and the nuclear contribution are the same for a given molecule, whether it is LERCCSD or an nLERCCSD calculation, the interplay between $AT_1$ and $T_1^\dag A T_1$ terms decide the importance of non-linear terms. For HgX, the $AT_1$ term in nLERCCSD calculations is always slightly larger in magnitude than the LERCCSD ones, while the net contributions from the $T_1^\dag A T_1$ terms, which are less significant, are the other way round. The resulting non-linear effects are not so important for SrF and BaF as seen in the earlier paragraph, while for HgX molecules, it becomes significant, with their effects changing the PDM by up to about 25 percent for HgI. 


We now conduct a survey of previous works on the effective electric fields and PDMs of the molecules that we have considered, in Table~\ref{tab:table1}. For the effective electric fields of BaF, we observe that effective core potential-restricted active space SCF (ECP-RASSCF) \cite{Kozlov} and restricted active space CI (RASCI) \cite{Nayak} methods give larger values, while the result from MRCI approach in Ref. \cite{Meyer2} estimate the values as being slightly lower, with respect to our nLERCCSD value. A discussion of the previous works on the effective electric fields of HgX and our improved estimate of the quantity using LERCCSD approach has already been presented in Ref.~\cite{HgX}, and hence we re-direct the reader to the earlier work. Our nLERCCSD results improve over the earlier LERCCSD and FFRCCSD results, as both of those were calculated using a DZ quality basis. Most of the works that calculate PDMs and which are not mentioned in the table, including Refs. \cite{Torring,Langhoff,Mestdagh,Allouche,Kobus}, have been expounded in our earlier works in detail~\cite{AEM,FFCC}, and therefore we only discuss in this paragraph the more recent works. The differences in the PDMs between the LERCCSD results in our earlier work and those in the present work for HgX are due to the choice of basis (DZ basis functions~\cite{HgX,AJP} in the former, as against a QZ basis in the current work). We observe that the values of PDM for SrF that are obtained by using complete active space self consistent field (CASSCF) approach to multi-reference CI (MRCI) and second-order Rayleigh-Schrodinger perturbation theory (RSPT2) \cite{Jardali} (which agrees with our nLERCCSD as well as FFRCCSD results) underestimate and overestimate the results with respect to experiment, respectively. The results for PDMs of SrF and BaF from Hao \textit{et al}~\cite{Hao} using exact two-component Hamiltonian-Fock Space coupled-cluster  (X2C-FSCC) formalism and the PDM of SrF from Sasmal \textit{et al}~\cite{Sasmal} using a relativistic Z-vector coupled-cluster approach (with both the works employing a QZ basis) agree closely with experimental values. However, we also note that while the Z-vector approach predicts the PDM of  SrF very accurately, it underestimates that of BaF~\cite{Talukdar}.  This existing difference in the PDMs of SrF and BaF between the nLERCCSD and the FFRCCSD approaches on one side and the Z-vector RCCSD approach on the other could possibly be resolved in future works that employ methods that are even more refined.  

\begin{figure}[t]
\centering
\includegraphics[width=10.0cm, height=7cm]{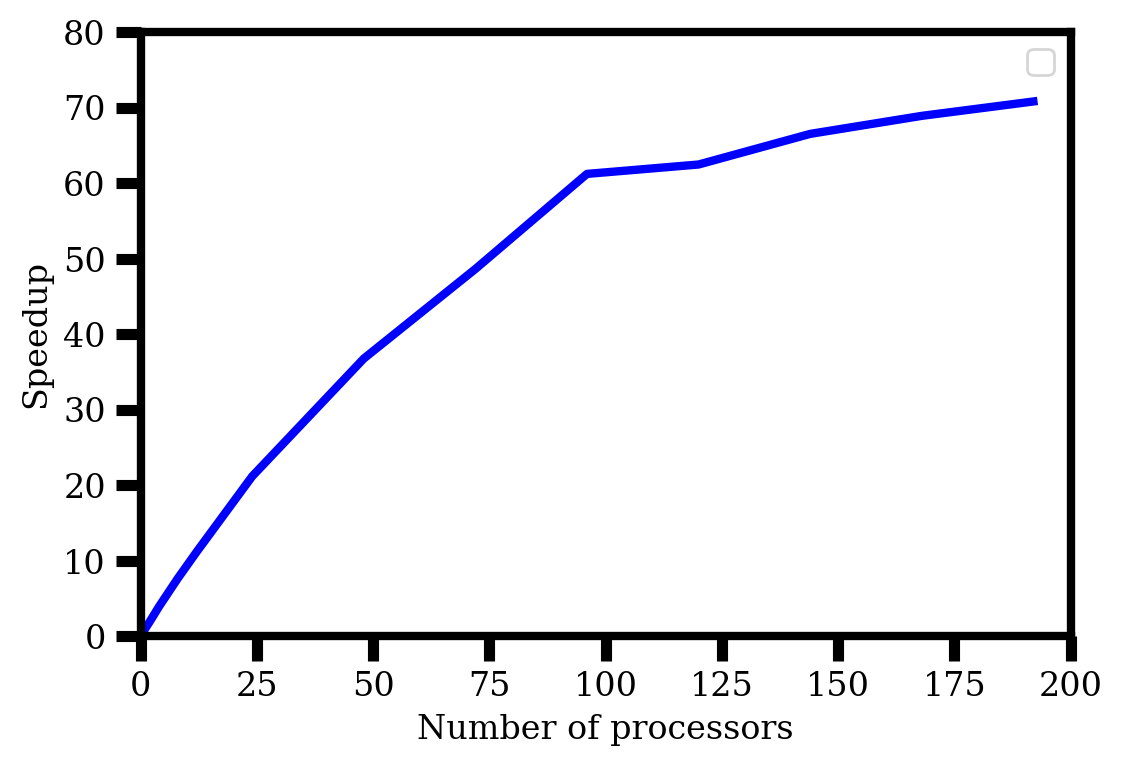} 
\caption{Plot showing the scaling behaviour of the program in the property evaluating expression for a representative system, SrF, with number of processors of our computer. The X-axis gives the number of processors, while the Y-axis is the speedup, $S_p = t_1/t_p$, where t is the time taken and the subscript denotes the number of processors. We have used a double-zeta quality basis for this purpose, and test up to 192 processors, as the parallelism in our code is limited by the number of virtual orbitals, which is 208 in this case.} 
\label{fig5}
\end{figure}

We now check for the scalability of our code that was parallelized using MPI. We do so by testing it with the SrF molecule, using a DZ basis. The code is to calculate both effective electric field and PDM of the molecule for this test. As the code is structured in a way that the extent of parallelization is limited by the number of virtuals, which is 208 in this case, we chose to study scaling up to 192 processors (across 8 nodes, and with 24 processors employed per node). The details of the computer (VIKRAM-100 super-computing facility at Physical Research Laboratory, Ahmedabad, India) that we used are: a 100-teraflop IBM nx360 M5 machine with 1848 processors. Each node has 24 processors (two Intel Xeon E5-2670 v3, each with 12 cores) and a memory of 256 GB RAM. The inter-process communication is via a 100\% non-blocking FAT Tree Topology FDR (56 Gbits/Sec) Infiniband network. We use an Intel 15.2 compiler, and impi5.0 and mkl libraries. As Figure~\ref{fig5} shows, our calculations indicate that the code is scalable up to this mark. In the figure, we plot the speedup against the number of processors, where the former is defined as $S_p = t_1/t_p$, with $t_p$ referring to the time taken for a computation with $p$ processors. Performing the computations in serial mode takes about 6.5 days, while calculations with 4 processors consumes around 2 days. The code takes  only 2.17 hours to finish with 192 processors. The walltime approaches saturation after 96 processors (2.51 hours to 2.17 hours from 96 to 192 processors), and hence the optimal number of processors to use is around 96, where we still get speed-up from 6.5 days to 2.51 hours. The walltimes are reliable as estimates, but not extremely accurate, as the computations were performed on a common cluster, and the speeds depend upon other factors such as the number of users, the computer's specifications, and type of jobs during the time interval across which our computations are done, although we took utmost care to ensure that no other application ran on the same node(s) as ours. However, our analysis is sufficient for the purposes of broadly demonstrating that our code is scalable to a reasonably large number of processors. 

We also estimate the errors in our calculations. We first examine the error due to choice of basis. We use QZ quality basis sets for our calculations, and as there is no 5-zeta basis that is available for us to carry out any kind of estimate, we calculate the effective electric fields and the PDMs at DZ level of basis with our nLERCCSD code. We find that the percentage fraction difference between DZ and QZ basis for $\mathcal{E}_\mathrm{eff}$ is around 3, 4, 5, and 7 percent for HgF, HgCl, HgBr, and HgI, respectively. We do not anticipate the difference between the DZ and QZ estimates to be over 10 percent for SrF and BaF either. Therefore, we do not expect that the difference between results from a higher quality basis set than QZ and those from a QZ basis should exceed 10 percent. Based on similar considerations, we estimate the error due to choice of basis for the PDM to be at most 15 percent. Next, we shall look at the errors due to the ignored non-linear terms. They are expected to be negligible, and we shall ascribe a conservative estimate of 2 percent, which is the percentage fraction difference between the DF values and the current nLERCCSD values for the HgX molecules. Lastly, we comment on the importance of triple and other higher excitations. Based on our previous works and error estimates in them, we expect that these excitations would be around 3 percent for the purposes of calculating $\mathcal{E}_\mathrm{eff}$~\cite{FFCC}, but could become important for PDMs. In conclusion, we linearly add the uncertainties and set an optimistic error estimate for the effective electric fields at about 15 percent. However, it is not so straightforward to set an error estimate for PDMs, as seen above, but we do not anticipate it to exceed 20 percent. 

\section{Conclusion}\label{sec4}

We have investigated contributions from the non-linear terms of the property evaluating expression of the relativistic coupled-cluster theory in the determination of permanent electric dipole moments and effective electric field due to electron electric dipole moment of SrF, BaF and mercury monohalides (HgX with X = F, Cl, Br, and I) molecules. We find that the inclusion of these terms at the singles and doubles excitations approximation brings the permanent electric dipole moments (PDMs) of SrF and BaF closer to the previously calculated finite-field relativistic coupled-cluster values, which were found to have overestimated the PDMs of the two molecules with respect to available measurements. The non-linear terms considerably change the PDMs of HgX systems. For all of the chosen molecules, the non-linear terms are found to not significantly change the values of effective electric fields with respect to the results from the linear expectation value approach. However, such a result is a consequence of several cancellations at work. Since accurate estimation of these quantities are of immense interest to probe new physics from the electron electric dipole moment studies using molecules, our analysis demonstrates importance of considering non-linear terms in relativistic coupled-cluster theory for their evaluations. We have also presented the scaling behaviour of our code with a representative SrF molecule, and discussed the error estimates.

\section*{Acknowledgments}

All the computations were performed on the VIKRAM-100 cluster at PRL, Ahmedabad.

\end{document}